\documentclass[10pt,a4paper]{article}
\usepackage[utf8]{inputenc}
\usepackage[T1]{fontenc}
\usepackage{amsmath}
\usepackage{amsfonts}
\usepackage{amssymb}
\usepackage{graphicx}
\usepackage{tabularx}
\usepackage{multirow}
\usepackage{caption}
\usepackage{float}
\usepackage[caption = false]{subfig}
\usepackage{hyperref}
\DeclareUnicodeCharacter{2212}{\textendash}
\author{Marcos Matabuena$^{1,*}$, Sherveen Riazati$^{2}$, 
Nick Caplan$^{3}$, Phil Hayes$^{3}$
 \\ $^{1}$CiTIUS (Centro Singular de Investigaci\'{o}n en Tecnolox\'{i}as Intelixentes), \\
	Universidade de Santiago de Compostela, Spain\\  $^{2}$Biomechanics, Rehabilitation and Integrative Neuroscience (BRaIN), \\ School of Medicine, University of Davis, United States\\
	$^{3}$Department of Sport and Exercise Sciences, School of Life Sciences,\\ Northumbria University,  Newcastle upon Tyne, United Kingdom \\
	$^{*}$\url{marcos.matabuena@usc.es}}

\title{Are Multilevel functional models the next step in sports biomechanics and wearable technology? A case study of Knee Biomechanics patterns in typical training sessions of recreational runners}

\begin{document}
	\maketitle
	
	\section*{abstract}
	This paper illustrates how multilevel functional models can detect and characterize biomechanical changes along different sport training sessions. Our analysis focuses on the relevant cases to identify differences in knee biomechanics in recreational runners during low and high-intensity exercise sessions with the same energy expenditure  by recording $20$ steps. To do so, we review the existing literature of multilevel models and then, we propose a new hypothesis test to look at the changes between different levels of the multilevel model as low and high-intensity training sessions. We also evaluate the reliability of measures recorded in three-dimension knee angles from the functional intra-class correlation coefficient (ICC) obtained from the decomposition performed with the multilevel funcional model taking into account $20$ measures recorded in each test. The results show that there are no statistically significant differences between the two modes of exercise. However, we have to be careful with the conclusions since, as we have shown, human gait-patterns are very individual and heterogeneous between groups of athletes, and other alternatives to the p-value may be more appropriate to detect statistical differences in biomechanical changes in this context.

	\section*{General overview and motivation}
 Advances in biosensors and digital medicine are improving disease monitoring and detection. A promising field for implementing these novel strategies is sports training and biomechanics. These tools  are a crucial element for optimizing athlete training and reducing the incidence of sports injuries. Multiple repeated measurements are collected from each individual over different sessions, weeks, or the entire season in these domains. So, it is essential to evaluate the changes produced along with relevant outcomes at different resolution levels scales among individuals. In addition, much of the information recorded is of a functional nature, such as the cycle of gait movement. Functional gait analysis enables a more accurate assessment of the effect of fatigue and the detection of potential injury risk factors. This paper illustrates how multilevel functional models can detect and characterize biomechanical changes along different training sessions. Besides, the multilevel models can provide a vectorial representation of different athlete training activities and feed supervised predictive models in various modeling tasks. Our analysis focuses on the relevant cases to identify differences in knee biomechanics in recreational runners during low and high-intensity exercise sessions with the same energy expenditure  by recording $20$ steps. To do so, we review the existing literature of multilevel models and then, we propose a new hypothesis test to look at the changes between different levels of the multilevel model as low and high-intensity training sessions. We also evaluate the reliability of measures recorded in three-dimension knee angles from the functional intra-class correlation coefficient (ICC) obtained from the decomposition performed with the multilevel funcional model taking into account $20$ measures recorded in each test. The results show that there are no statistically significant differences between the two modes of exercise. However, we have to be careful with the conclusions since, as we have shown, human gait-patterns are very individual and heterogeneous between groups of athletes, and other alternatives to the p-value may be more appropriate to detect statistical differences in biomechanical changes in this context.  
	
	\section{Introduction}


In recent years, there has been a big increase in the availability of powerful biosensors. These are now capable of monitoring an individual’s energy expenditure with great accuracy and measuring various physiological and biomechanical variables in real-time. This provides the opportunity to have a unique assessment of an individual’s physical capability \cite{lencioni2019human} and performance and thus, be able to schedule optimal interventions over time \cite{kosorok2019precision, buford2013toward}. One field that can benefit from the intensive use of these technologies is biomechanics \cite{ibrahim2021artificial,uhlrich2020personalization}. In both sports and general populations, abnormal movement patterns are synonymous with muscular and motor problems, risk of injury, or even the appearance of severe neurological diseases such as Parkinson’s \cite{morris2001biomechanics}. Therefore, the detection and characterization of these abnormal movement patterns in biological activities such as walking and running, are essential in areas beyond professional and sports medicine, such as clinical medicine \cite{chia2020decision}.

Nowadays, with the growing boom of wearables, these technologies are being democratized, and their use is increasingly more common among the general population, such as amateur runners. In this setting, the remote control of athlete training and even monitoring their daily routine out of sport activity is feasible. Although we are in the early stages of this technological revolution, the first research papers are appearing, which through high-resolution data gathered with biosensors, can begin to answer unknown and complex  questions about the relationship between training load \cite{cardinale2017wearable}, daily biomechanical patterns \cite{10.1093/biostatistics/kxz033}, and injury prediction \cite{bittencourt2016complex, malone2017unpacking}. Furthermore, they may even enable us to build predictive models that support decision-making and help optimize the performance \cite{matabuena2019improved,hemingway2020narrative, piatrikova2021monitoring}.   For example, several contemporary works provide new epidemiological knowledge using  biomechanical data of human locomotion  \cite{10.1093/biostatistics/kxz033}, \cite{WARMENHOVEN2020110106}. Other papers have tried to predict sports injuries\cite{rossi2018effective} or other motor or neurological diseases prematurely \cite{belic2019artificial}, or even the impact of therapy together with their prognosis in the recovery phase after surgery \cite{karas2020predicting}.

The rising proliferation of running as a sporting activity carries a substantial risk for recreational runners who often perform high-intensity training as interval sessions without a formal training schedule. For recreational runners there has been an increased prevalence of running related injures the most common of which is the knee \cite{van2007incidence,messier2008risk}.

To date, several works have studied the aetiology of running related knee injuries in recreational runners, some even using  $3$-dimensional analysis \cite{messier2008risk}. However, to the best of our knowledge, no studies have compared biomechanical changes during high intensity interval training (HIIT) compared to lower intensity continuous running. Moreover,  some essential questions remain unanswered, for example the reliability of biomechanical measures at the knee in two or more HIIT training sessions.

Traditionally, gait analysis has been performed at fixed points within the gait cycle. A more detailed and meaningful analysis can be attained by using a complete stride cycle with functional data analysis (FDA) \cite{febrero2012statistical} techniques. These analyses can provide a greater insight into understanding the mechanics of locomotion especially as a runner fatigues. Under different fatigue conditions it would be possible to identify with greater clarity, changes that take place within each part of the gait cycle.

The predominant data-analysis practice in biomechanics is to summarize the curve recorded for each stride using several statistical metrics and applied standard multivariate techniques, although  there is a loss of information with this approach. There is a rising popularity of using functional data in biomechanics, with the purpose too bridge this gap between complex statistical modeling with functional data and standard analysis between practitioners. An interesting paper \cite{WARMENHOVEN2020110106} explains a global target audience using principal functional component analysis to more accurately analyze biomechanics data. However, this methodology does not take into account that we can obtain multiple strides per individual in each training session.

Here, the general procedure is to normalize the curve obtained for step and body segment to the $[0,1]$ interval and take the mean of the different curves recorded and create an average functional curve for analysis. Nevertheless, this procedure can be suboptimal because the constructed mean representation ignores the individual variability between the distinct steps of the same individual, something crucial in the evaluation of the movement patterns in some settings. In addition, the mean curve can be a summary measure of the information that is very sensitive to outliers, something that is frequent in biomechanical data. This is particularly true in the measurement of movements performed at high or low speed, where sensor and/or human variability often increase. Moreover, we often need to compare the effect of an intervention along the different training sessions on different days, and for this, we have several repeated measures per individual in different periods. For all this, a more natural analysis is to exploit the advantages of multi-level functional models that allow the analysis of several hierarchy levels. With these models, we can incorporate  in a natural way, into the same statistical models, biomechanical patterns using a significant fraction of training session, test or event data from a complete session. These methods also allow the capture  the  variations in different periods at a intra-inter individual level.

Surprisingly, there is little use of these FDA  techniques within the literature, either in sport or indeed other clinical areas \cite{ullah2013applications}. Both of these areas would benefit from larger data sets, be that longitudinally or more strides or conditions.

The objective of this paper is two-fold. First, we will introduce the analysis of multi-level FDA from the methodological point of view. After, we will illustrate from an applied point of view that these biomechanical methods analyze several exciting research questions. For this purpose, we use a  sample composed of $19$ athletes during two different training sessions, one moderate and one high speed, in a controlled laboratory environment. During these training sessions we measured  knee patterns with a tridimensional sensor of $20$ strides during the stance phase.

The structure of the paper is as follows. First, we introduce multi-level models with FDA, as review the literature. We then describe the study sample; we do a more in-depth analysis of tridimensional changes to acquire new biomechanic knowledge. 	Finally, we discuss the results and future challenges in multi-level models in biomechanics and other sport biosensor data.

Given the double target audience of this paper, to maintain the interest for biomechanical practitioners that do not have a specific interest in the mathematical details, we a illustrate multi-level methodology to show  biomechanical applications to our data analysis example:  

\begin{itemize}
	\item What are the correlations between knee functional running patterns during a HIIT training session and the loss of force production in training?

	\item What is the reliability of the  functional running parameters in two independent HIIT training seasons?
	
	\item 	Are there differences in knee angles between a continuous running session and HIIT training?
	
	\item Are functional biomechanics patterns very individual between runners? 
	
	\item Is it appropriate to use p-value to detect biomechanical changes in the practice?

\end{itemize}

	\section{ Multi-level functional data analysis}

	\subsection{General overview of state-of-art of multi-level models}

	Functional data analysis \cite{febrero2012statistical, cuevas2014partial, wang2016functional} with a multi-level structure and repeated measurements is a field that has received substantial attention in recent years (see for example \cite{lee2018bayesian,li2020regression}). In the statistical community, it appeared in the literature as an essential new methodology to statistical practitioners. These techniques have been applied successfully to answer central scientific questions in such heterogeneous domains, to study the variability between subjects, days, tests, physical activity patterns,  speech, or sleep quality monitoring  \cite{xiao2015quantifying,huang2019multilevel,park2018simple,martinez2013study, pouplier2017mixed, di2009multilevel}. Probably the first work that addressed mixed functional data problem was back in $2003$—in this work \cite{morris2003wavelet}, using wavelets and a Bayesian estimation procedure, the effect of type f dietary fat on O6-methylguanine-DNA-methyltransferase (MGMT), an important biomarker in early colon carcinogenesis, was explored. Since then, different models have appeared in the functional data analysis literature that have modeled different hierarchy levels, including nested and crossed structures—using diverse estimation strategies adapted to the nature of the real problem that the authors treat to solve. From a general point of view, the different data characteristics involved in the creation of new mixed functional data models are the number of data recorded \cite{zipunnikov2011multilevel},	 the density of the functional data \cite{https://doi.org/10.1002/sta4.50},
	 the number of replicates in each unit of the hierarchical structure \cite{zipunnikov2011multilevel},	  the structure of dependence between levels of the hierarchy or replicates \cite{10.1093/biostatistics/kxp058, staicu2012modeling}, or the dependence between covariates in multidimensional functional problems \cite{volkmann2021multivariate}. For example, in some relevant applications, such as the analysis of longitudinal data obtained from medical images using nuclear magnetic resonance, the high computational demands of the estimation of image correlation operators and the calculation of projections between subjects and visits have resulted in a series of papers using new computationally efficient multi-level methods that scale well in problems involving millions of data \cite{zipunnikov2011multilevel, zipunnikov2014longitudinal}. In a similar way, essential progress has been made in recent years in the smoothing of correlation operators, whereby at the moment, in problems with half a million covariates, the smoothing can be done in a few seconds \cite{xiao2016fast, cederbaum2018fast}.  In the reverse situation with sparse functional data, efficient estimation methods have also been proposed \cite{https://doi.org/10.1002/sta4.50, xiao2018fast, li2020fast}, but from a statistical perspective, data variability increases due to the low density of functional data in this framework, and the estimation problems are magnified. Using this approach, several works have proposed different inferential contributions such as re-sampling bootstrap methods to perform inferential tasks such as calculating confidence intervals or comparing the equality of means between groups of subjects by exploiting the rich source of information several measures of the same individual in biological problems \cite{crainiceanu2012bootstrap, goldsmith2013corrected, park2018simple}. Also, some of the previous multi-level models have been generalized to introduce the impact of specific covariates on the levels of variability of the different levels of hierarchy and subjects so in supervised and unsupervised problems \cite{crainiceanu2009generalized, 10.1093/biostatistics/kxs051,xiao2015quantifying,scheipl2015functional}. Furthermore, new methods have been proposed in a more general set up as other complex objects such as functional matrix structure data \cite{huang2019multilevel}. This technique, has facilitated opportunities to study the variations of physical activity patterns in a group of subjects with cardiac pathology over several weeks, highlighting the power of these models to solve real complex problems.

	\subsection{Mathematical models}
	
	\subsubsection{Mathematical foundations of standard functional principal component analysis}
	
	Let $X(t),$ $t\in [0,1]$, be a random function with mean $\mu(t)= E(X(t))$ and covariance function $\Sigma(t,s)= E((X(t)-\mu(t)(X(s)-\mu(s))$ for all $t,s$ $\in$ $[0,1]$. The heart of many functional data analysis models is based on calculating modes of variability of the random function $X(t)$ based on the spectral decomposition of the covariance operator $\Sigma(\cdot,\cdot)$ in a set of eigen-functions $\{e_i(\cdot)\}^{\infty}_{i=1}$ and eigenvalues $\{\lambda_{i}\}^{\infty}_{i=1}$, and where, we suppose that $\lambda_{1}\geq \lambda_{2} \geq \cdots $. Thus, from the decomposition of Karhunen-Loève we have 
	
	\begin{equation}
	X(t)= \mu(t)+\sum_{k=1}^{\infty} c_k e_k(t)
	\end{equation}
	
	where, $c_k= \int_{0}^{1}(X(t)-\mu(t)) e_k(t) dt$ being $c_k$'s incorrelated random variables with mean  zero, and variance $\lambda_{k}$. These variables are usually known as scores or loading variables.

	In the real-world setting, we have $n$ realizations generally independent of the process  $X(\cdot)$, $X^{1}(\cdot)$, $\dots$,  $X^{n}(\cdot)$, but only we observe a sample of  $n$ vectors of length $m$, $X^1$, $\dots$, $X^n$, in a grid $\{0\leq t_1< \dots <t_{m}\leq 1\}$, and  where $X_{j}^i= X^{i}(t_j)$ for all $i=1,\dots,n$, $j=1,\dots,m$. Next, by simplicity, to refer  $X_{j}^i$, we use $X^{i}(t_j)$.

	The simpler estimator of $\Sigma$ is,

	\begin{equation}
	\hat{\Sigma}(t_j,t_k)=  \frac{1}{n} \sum_{i=1}^{n}(X^i(t_j) -\hat{\mu}(t_j))(X^i(t_k) -\hat{\mu}(t_k)),
	\end{equation}
	
	where $\hat{\mu}(t_j)=  \frac{1}{n}\sum_{i=1}^{n} X_i(t_j)$, for all $j=1,\dots, m$.

	A major step here, in many applications where observations are subjected to a large measurement error, is the smoothing process, to ensure the optimal performance of the empirical estimator $\hat{\Sigma}$. Three different strategies have generally been used in the literature \cite{shang2014survey, cederbaum2018fast}:
i)	Smoothing of the original functional data; ii)
	Introduction of a regularization term in the estimation of $\hat{\Sigma}$; and iii) direct application of a smoothing procedure  in the raw estimation of $\hat{\Sigma}$.	Subsequently, we denote by $\hat{\Sigma}_{Smooth}$ the smoothed version of $\hat{\Sigma}$ by any of the three previous procedures.

	The next step is to calculate the auto-vectors and auto-values of $\hat{\Sigma}_{Smooth}$, according to the spectral theory of linear algebra, as happens in the classical context of principal component analysis in multivariate statistics. After performing this procedure, and selecting the first $K<m$ auto-vectors $\{\hat{e}^i\}_{i=1}^{K}$ and auto-values $\{\hat{\lambda}_i\}_{i=1}^{K}$ we obtain the following decomposition:
	
	\begin{equation}
	X^{i}(t_j) \approx \hat{\mu}(t_j) + \sum_{k=1}^{K} \hat{c}^{k} \hat{e}^{i}_k  \hspace{0.25cm}    (i=1,\cdots,n; j=1,\cdots, m),
	\end{equation}	
	
	where $\hat{c}^{i}= \langle X^{i}-\hat{\mu},\hat{e}^{i}\rangle $, denoting with $\langle,\rangle$ the usual scalar-product and being $\hat{e}^{i}_j$ the $j$ component of the auto-vector $\hat{e}^i$.
	
	More details of this procedure can be found in the following reviews and general books of functional data analysis \cite{horvath2012inference,shang2014survey, kokoszka2017introduction}, where different estimation procedures of the number $K$, of components are established \cite{li2013selecting}. For more theoretical aspects of the estimators such as asymptotic properties we refer the reader to \cite{hall2006properties}.

	\subsubsection{Introduction of functional multilevel  models}\label{sec:multilevel1}

	In the previous Section, we have seen how to carry out a principal component analysis when we observed $n$ independent functional data.  In practice, in biomedical and sports applications, when patients or athletes are analyzed at different moments in time, for example, the training load  throughout a season, it is common to have several repeated and correlated measurements. Therefore, the previous procedure may be inadequate. Before starting, we introduce some extra notation to describe this scenario.

	Le $X^{i,j}(t)$, $t\in [0,1]$, the functional datum of individual $i$, with the measure $j$, for $i=1,\cdots, n$, $j=1,\cdots, n_i$, that for simplicity we assume the method introduces that  $n_i= J$ (for all $i=1,\cdots, n$).  
	
	To start, let-consider the two-way functional ANOVA model, whose structure is introduced below:
	
	\begin{equation}{\label{sec:multilevel}}
	X^{i,j}(t)= \mu(t)+\nu^{j}(t)+ Z^{i}(t)+W^{i,j}(t)   \hspace{0.2cm} (i=1,\dots, n; j=1,\dots, J),  
	\end{equation}
	
	where $\mu(t)$ is the mean global, $\nu^{j}(t)$, is the mean of measure  $j$, $Z^{i}(t)$   is the subject-specific deviation from the measure-specific mean function, and $W^{i,j}(t)$ is the residual subject- and measure-specific deviation from the subject-specific mean. In this framework,  $\mu(t)$ and $\nu^{j}(t)$  are treated as fixed functions, while $Z^i(t)$ and $W^{i,j}(t)$  are treated as    random function of mean zero. Moreover, with the proposal of identification correctly the model, we assume that $Z^{i}(t)$, and  $W^{i,j}(t)$ are random uncorrelated functions. In many applications, we note $\nu^{j}(t)$  could be set to  zero when functional responses are interchangeable within different measures, and the model becomes a one-way functional ANOVA.

	In the literature of multi-level models, the functions $Z^{i}(t)$'s are known as the $1$-level of functions, while $W^{i,j}(t)$'s functions composed the $2$-level.

	Again, the foremost step in a multi-level functional component analysis model is to rely on the Karhunen-Loève decomposition. For example, in the model defined by the equation \ref{sec:multilevel}, we have
	
	\begin{equation}{\label{sec:multilevel2}}
	Z^{i}(t)= \sum_{k=1}^{\infty} c_{k}^{i} e_{k}^{(1)}(t)   \hspace{0.4cm} W^{i,j}(t)=  \sum_{k=1}^{\infty} d_{k}^{i,j} e^{(2)}_{k}(t) 
	\end{equation}

	where $\{e_{k}^{(1)}\}^{\infty}_{k=1}$ y $\{e_{k}^{(2)}\}^{\infty}_{k=1}$ are the auto-functions related to the random functions of the levels $1$, $2$, respectively, while $\{c_{k}^{i}\}^{\infty}_{k=1}$, $\{d_{k}^{i,j}\}^{\infty}_{k=1}$ are the scores o loading variables for all $i=1,\dots, n$; $j=1,\dots, J$. 
	
	In more compact form, the equation \ref{sec:multilevel}, is rewritten as

	\begin{equation}{\label{sec:multilevel3}}
	X^{i,j}(t)= \mu(t)+\nu^{j}(t)+\sum_{k=1}^{\infty} c_{k}^{i} e_{k}^{(1)}(t)+\sum_{k=1}^{\infty} d_{k}^{i,j} e^{(2)}_{k}(t)   \hspace{0.2cm} (i=1,\cdots, n; j=1,\cdots, J).  
	\end{equation}

	Importantly, in this models, the functions $\{e_{k}^{(1)}\}^{\infty}_{k=1}$ $\{e_{k}^{(2)}\}^{\infty}_{k=1}$ are ortho-normal basis in the space of square functions, but in general the functions that compose each function bases are not orthogonal with each other, which implies that the estimation of scores is not simple in practice, a topic that we discuss later. Moreover, score`s  $\{c_{k}^{i}\}^{\infty}_{k=1}$ and  $\{d_{k}^{i,j}\}^{\infty}_{k=1}$ are random variables of mean zero and with variance are given by covariance funtions of stochastic process   $Z^i(t)$'s and  $W^{i,j}(t)$'s.
	
	Below, we explain how to calculate the auto-functions and auto-values of the model defined in the equation  \ref{sec:multilevel3}. Let $\Sigma_{T}(s,t)=  Cov(X^{i,j}(s), X^{i,j}(t))$ be the overall covariance function and $\Sigma_{B}(s,t)=  Cov(X^{i,j}(s), X^{i,k}(t))$ the covariance function between the units of the second level setting the effect of the first level. Applying Mercer's theorem, we can see that it verifies 
	$\Sigma_{T}(s,t)= \sum_{k=1}^{\infty} e_{k}^{(1)}(s) e_{k}^{(1)}(t) \lambda_{k}^{(1)}+\sum_{k=1}^{\infty} e_{k}^{(2)}(s) e_{k}^{(2)}(t) \lambda_{k}^{(2)}$ and $\Sigma_{B}(s,t)= \sum_{k=1}^{\infty} e_{k}^{(1)}(s) e_{k}^{(1)}(t) \lambda_{k}^{(1)}$. Defining  $\Sigma_{W}(s,t)=  \Sigma_{T}(s,t)= \Sigma_{T}(s,t)-\Sigma_{B}(s,t)=\sum_{k=1}^{\infty} e_{k}^{(2)}(s) e_{k}^{(2)}(t) \lambda_{k}^{(2)}$,  where the indices $T$,$B$, and $W$ are used to refer to the "total," "between," and "within" subject covariances. 
	
	As in the previous Section, the curves are observed uniquely, in a grid of points $\{0\leq t_1< \cdots <t_{m}\leq 1\}$ and, in this situation we have to perform the empirical estimators $\hat{\Sigma}_{T}(t_s,t_k)$, $\hat{\Sigma}_{W}(t_s,t_k)$, $\hat{\Sigma}_{B}(t_s,t_k)$. Unlike in the preceding Section, we only observe directly  information from the process $X^{i,j}(t)$, and, it is possible to estimate the covariance matrix, $\hat{\Sigma}_{T}(t_s,t_k)$, according the  usual empirical estimator, that is,

	\begin{equation}
	\hat{\Sigma}_{T}(t_s,t_k)= \frac{1}{nJ} \sum_{i=1}^{n}\sum_{j=1}^{J}(X^{i,j}(t_s)-\hat{\mu}(t_s)-\hat{\eta}^{j}(t_s))(X^{i,j}(t_k)-\hat{\mu}(t_k)-\hat{\eta}^{j}(t_k)).
	\end{equation}

	To estimate the covariance operator $\Sigma_{B}(s,t)$, it is enough to appeal to the method of moments or in an equivalent way to the covariance estimator through a $U$-statistic estimator,  
	
	\begin{equation}
	\hat{\Sigma}_{B}(t_s,t_k)= \frac{1}{nJ(J-1)} \sum_{i=1}^{n}\sum_{j=1}^{J} \sum_{j^{\prime}  \neq j}^{J}   (X^{i,j}(t_s)-\hat{\mu}(t_s)-\hat{\eta}^{j}(t_s))(X^{i,j^{\prime}} (t_k)-\hat{\mu}(t_k)-\hat{\eta}^{j^{\prime}`}(t_k)).
	\end{equation}

	As $\hat{\Sigma}_{W}= \hat{\Sigma}_{T}-\hat{\Sigma}_{B}$, is not necessarily a defined positive matrix in the sample context, we have to trim the eigenvalue-eigenvector  pair where the eigenvalue is negative.

	Finally, for the calculation of the scores $\{c_{k}^{i}\}^{\infty}_{k=1}$ y  $\{d_{k}^{i,j}\}^{\infty}_{k=1}$ (for all $i=1,\dots, n$; $j=1,\dots, J$). Different estimation strategies can be used. We highlight computationally intensive methods such as Markov chain Monte Carlo (MCMC) \cite{di2009multilevel},  projection algorithms designed for this problem, or the most computationally efficient and used in practice method: the best linear unbiased prediction estimator for mixed models (BLUP) \cite{robinson1991blup}, which is detailed for the multilevel problem defined in this Section, in the following reference \cite{https://doi.org/10.1002/sta4.50}.

	\subsubsection{More general extensions}\label{sec:multilevel2}
	
	Different levels of hierarchy may appear in real problems that can be nested as in the previous Section or crossed. Following \cite{shou2015structured},  the different situations that usually occur are listed in Table \ref{table:sec1}. All these models have the same structure $X(t)= \mu(t)+ \sum latent processes +\epsilon(t)$, where $\mu(t)$ is the mean curve or fixed effect and $\epsilon(t)$ is a white noise, $\epsilon(t) \sim N(0,\sigma^2)$ for all $t\in [0,1]$.  The latent processes are assumed to be zero-mean and square-integrable so that they are identifiable, and the standard statistical assumptions for scalar outcomes can be generalized to functional data. In this way, the total variability of a functional outcome is decomposed into a sum of process-specific variations plus $\sigma^{2}$. 
	
	Both nested and crossover models can be used to employ a general estimation strategies. Below  we  summarize the steps necessary to do so, which are analogous to those explained in the previous Section: 
	
	\begin{enumerate}
		
		\item Estimate the means  and covariance functions involved in the differents models via moment methods.
		
		\item With the estimated covariance functions, calculate an appropriate number of $K$- auto-values and auto-vectors along the different levels of hierarchy that collect the different modes of variability in a precise way so that the problem we want to address.
		
		\item Estimate the scores using the BLUP estimator  \cite{robinson1991blup}, as it is done in \cite{shou2015structured} based on \cite{zipunnikov2011multilevel} and \cite{crainiceanu2009generalized}.

	\end{enumerate}

	\begin{center}

		\begin{table}[ht!]
			\scalebox{0.8}{
			
			\begin{tabular}{ |c|c|c|c| }

				\hline
				& Model & Structure \\
				\hline
				\multirow{3}{4em}{Nested} & (N1) One-way & $X^{i}(t)= \mu(t)+Z^{i}(t)+\epsilon_{i}(t)     $ \\ 
				& (N2) Two-way & $X^{i,j}(t)= \mu(t)+Z^{i}(t)+W^{i,j}+  \epsilon_{i,j}(t)    $ \\ 
				& (N3) Three-way & $X^{i,j,k}(t)=\mu(t)+Z^{i}(t)+W^{i,j}+ U^{i,j,k}+ \epsilon_{i,j,k}(t) $ \\
				& (NM) Multi-way & $X^{i_1,i_2,\cdots, i_r}(t)=\mu(t)+R_{(1)}^{i_1}(t)+R_{(2)}^{i_2}+\cdots R_{(r)}^{i_r}+ \epsilon_{i_1,i_2,\cdots, i_r}(t) $ \\	
				\hline
				\multirow{3}{4em}{Crossed} & (C2) Two-way & $X^{i,j}(t)= \mu(t)+ \eta^{j}(t)+Z^{i}(t)+W^{i,j}+  \epsilon_{i,j}(t)    $ \\ 
				& (C2s) Two-way sub & $X^{i,j,k}(t)=\mu(t)+\eta^{j}+Z^{i}(t)+W^{i,j}+ U^{i,j,k}+ \epsilon_{i,j,k}(t) $ \\
				& (CM) Multi-way & $X^{i_1,i_2,\dots, i_ru}(t)=\mu(t)+R^{S_i}(t)+R^{S_2}+\cdots R^{S_r}+ \epsilon_{i_1,i_2,\dots, i_ru}(t) $ \\
				\hline
			\end{tabular}}
			\label{table:tabla1}\caption{
				Structured functional models.  For nested models, $i=1,2,\dots,n$;$j=1,2,\dots,n_i$; $k=1,2,\dots,K_{ij}$; $i_1=1,2,\dots,I_1$, $i_2=1,2,\dots, I_{2i1},\dots, i_r=1,2,\dots, i_{r_{i_1},i_2,\dots,i_{r_1}}$. For crossed designs, $i=1,2, \dots,n;j=1,2,\cdots,J;k=1,2,\cdots,n_{ij}$;  (C2s) "Two-way sub" stands for "Two-way crossed design with subsampling"; (CM) contains combinations of anys $(s=1,2,\dots,r)$ subset of the latent processes, as well as repeated measurements within each cell. $S_1,S_2,\dots,S_d$ $\in\{i_{k_1}i_{k_2} \dots,   i_{k_s}, u:k_1,k_2,\dots,k_s \in (1,2, \dots,r), u\in(\emptyset,1,2,\dots,I_{i_1i_2,\dots,i_r}), s \leq r\}$, $u$ is the index for repeated observation in cell $(i_{k1},i_{k2},\dots, i_{kr})$. $\epsilon(t)$ is a random errror $N(0,\sigma^{2})$} \label{table:sec1}
			
		\end{table}
		
	\end{center}

	In the step $1$, to estimate the covariance functions in the different models mentioned above, a general estimation strategy proposed in \cite{koch1968some} can be used. For example, following the notation and problem $(N2)$ defined in the previous Section, $\hat{\Sigma}_{T}$, $\hat{\Sigma}_W$, $\hat{\Sigma}_B$, can be expressed with the following sandwich structure:

	\begin{equation}\label{eqn:sand}
	\hat{\Sigma}_{T}= X G_{T} X^{T}   \hspace{0.2cm} \hat{\Sigma}_{W}= X G_{W} X^{T}  \hspace{0.2cm} \hat{\Sigma}_{B}= X G_{B} X^{T},
	\end{equation}

	where $X$ is a matrix of size $(nJ)\times m$ that records the different observations of all the individuals and levels of hierarchy, while $G_{T}$, $G_{W}$ and $G_{B}$ are design-specific matrices  of dimension $m\times m$.
	
	In particular, the usual co-variance matrix $\hat{\Sigma}_{T}$ is written as $\hat{\Sigma}_{T}= X G_{T} X^{T}$, where $G_{T}= \frac{1}{nJ} (I-11^{T})$ where $I$ denotes the identity matrix, and with $1$, we denote  $m$ length vector with all ones. More details about these procedures, as well as about the selection of the components and the score estimation, can be found in the following references \cite{di2009multilevel,shou2015structured}.

	\subsubsection{Intra-class correlation coefficient (ICC)}

	A significant problem when several repeated measurements are collected from a subject over different days or other periods is to determine how much variability is explained by the subjects' effect and how much by making different measurements over different levels of the hierarchy. This issue is in the literature and is known as the process of estimating the coefficient of intra-class correlation (ICC) \cite{muller1994critical} that pursues to estimate the variability of measuring a subject in conditions that are assumed to be standardized across different tests. The estimation of ICC is crucial, for example, in the field of clinical laboratory testing, where we want to use clinical variables for the monitoring and diagnosis of patients that are not modified abruptly between days by a problem of error of measurement of the device, and by the intra-day variability of individuals, see an example of the above in diabetes in \cite{Selvin2007short}. In biomechanics and exercise sciences, the ICC's quantification is also critical in searching for objective criterium  to assess performance and control the individual's degree of fatigue \cite{van2002reliability,koldenhoven2018validation}. Although a variable may have a high variability, it can be a very useful criterion  for decision-making. In this case, it is necessary to make several measurements to capture that variable accurately. The ICC can also quantify how many measures we have to make to capture variable distribution with enough accuracy.

	The first model where the ICC was estimated is $(N2)$ of the Table \ref{table:sec1}. In this scenario, we have

	\begin{equation}
	X^{i,j}(t)= \mu(t)+Z^{i}(t)+ W^{i,j}+  \epsilon_{i}(t).   
	\end{equation}

	Fixed  $t\in [0,1]$, by analogy with a univariate non-functional case, the proportion of the total explained variable by the effect of the subjects at that point, is given by

\begin{equation}
\rho(t)= \frac{Var(Z^{i}(t))}{Var(Z^{i}(t) +W^{i,j}(t)+\epsilon_{i}(t))}, 
\end{equation}

being $\rho(t)$,  the intra-class correlation coefficient in the point, $t$.

In a straightforward way, the ICC can be generalized as a global measure at the functional level, see for example (\cite{shou2013quantifying}) comparing the total variability collected by the involved co-variance operators with the variability modes $K_{A}$, and $K_{W}$ and the global white noise $\epsilon(t)$ (covariance functions at the levels $1$ and $2$). Thus, the global ICC, denoted as $\rho$, is 

\begin{equation}
\rho= \frac{tr(K_{A})}{tr(K_{A})+ tr(K_{W})+\sigma^{2}}= \frac{\sum_{k=1}^{\infty} \lambda_{k}^{(1)}}{\sum_{k=1}^{\infty} \lambda_{k}^{(1)}+ \sum_{k=1}^{\infty} \lambda_{k}^{(2)} + \sigma^{2} }, 
\end{equation}

wherewith $tr(\cdot)$, we denote the trace operation, and where we are using the notation of the Section \ref{sec:multilevel1}. We have to point out that the homoscedastic error-term $\epsilon(t)$ has been included according to the convention followed in the Section \ref{sec:multilevel2}-a more general setting that source of random variable is decomposed into an independent term.

The ICC can be calculated in more complex multi-level models. Suppose that we wish to use model (N3) and then, we have three levels. To develop such a task, it is enough to divide the source of variability generated by the hierarchy associated with subjects by all variability sources, that is:

\begin{equation}
\rho=  \frac{\sum_{k=1}^{\infty} \lambda_{k}^{(1)}}{\sum_{k=1}^{\infty} \lambda_{k}^{(1)}+ \sum_{k=1}^{\infty} \lambda_{k}^{(2)}+\sum_{k=1}^{\infty} \lambda_{k}^{(3)}  + \sigma^{2}}. 
\end{equation}

Recently, the intra-class correlation coefficient has been extended for objects that live in complex spaces where
similarity between objects can be computed by the particular distance
 \cite{xu2020generalized}.  	
	
	
		\subsubsection{Hypothesis testing between different levels}

	Consider the model (N3) specified on Table \ref{table:sec1} :

	\begin{equation}
	X^{i,j,k}(t)=\mu(t)+Z^{i}(t)+W^{i,j}(t)+ U^{i,j,k}(t)+ \epsilon_{i,j,k}(t), 
	\end{equation}
	
	where $i=1,2,\dots,n;j= 1,2,\dots,n_i;k=1,2,\dots,K_{ij}$.

	Without a loss of generality, suppose that the first level is the individual, the second is the test performed (HIIT or CTR), and the last level is the stride number. Comparisons between the differences in HIIT run and CTR are of considerable interest in biomechanical studies. To do this, we need to compare the difference between $2-$levels effect functions
	
	\begin{equation}
	    W^{i, HIT}(t) \hspace{0.2cm}
	    \text{and} \hspace{0.2cm}
	    W^{i, CTR}(t), \hspace{0.2cm} \forall t\in [0,1], i=1,\dots,n. 
	\end{equation}

Evoking again, Karhunen-Loève's decomposition,	we know that $W^{i, HIT}(t)\approx \sum_{k=1}^{m}    d^{i,HIT}_{k} e^{(2)}_k (t)$ and  $W^{i, CTR}(t)\approx \sum_{k=1}^{m}  d^{i,CTR}_{k} e^{(2)}_k (t)$. Then, to test the null hypothesis in a distributional sense  $H_0: W^{HIT}= W^{CTR}$, we can test the  score values in a distribution,  as follows:

	\begin{equation*}
	d_k^{HIT} \overset{D}{=} d_{k}^{CTR} \hspace{0.2cm} (k=1,\dots, m)	,\end{equation*}
	
	and it is expected that as $m,n\to \infty$, we have asymptotic test consistency.

	In practice, fixed $k$,  we can test univariate distribution changes with the estimated score of the second level composed of HIIT and CRT runs effects respectively, $\{\hat{d}^{i,HIT}_{k}\}^{n}_{i=1}$, $\{\hat{d}^{i,CRT}_{k}\}^{n}_{i=1}$. For this purpose, we can use the rich family test that provides energy distance methodology \cite{rizzo2016energy}, or with classical tests such as Kolmogorov-Smirnov or Crammer-Von-Misses. As we applied univariate-test $m$ times and obtained $m$ marginal p-values (for each score), we must apply false discovery rate \cite{benjamini1995controlling} or other criteriums to performer corrections for multiple comparisons to control type $1$ error under the null hypothesis. Finally, we return as global p-value $\min\{p^{*}_1,\cdots, p^{*}_m\}$, where $p^{*}_{i}$ denotes the adjusted p-value for score $i$. A similar methodology introduced here, was used in the standard set-up of hypothesis testing with functional data \cite{pomann2013two} out of a multilevel data framework.

	\subsection{Summary of functional multi-level models. What is the reason that this models are so important?}
	
	The increasing ability to store different profiles and functions of different variables that measure individuals' health from a broad spectrum of perspectives at different time scales provides several methodological challenges of statistical analysis that multilevel models can solve. In particular:

	\begin{itemize}
		\item  We can obtain a vectorial representation for each individual that captures the differences between individuals in a context of repeated and longitudinal measures. 
		\item We can obtain the same representation for each individual in different hierarchical levels, for example, in a specific run and specific step recorded.
		
		\item For a specific individual, we can estimate the differences between different hierarchical levels. In addition, we can quantify intra and inter individuals' variability in all model levels. With this model, we can see under specific conditions, the specific modes of variability and compare with other conditions.
		
		\item We can obtain reliability measures as ICC or compare through hypothesis testing changes along a group of individuals or test conditions with paired and repeated measures.  We can do this with the methodology previously established or following \cite{crainiceanu2012bootstrap}.
		
	\end{itemize}

	\section{Biomechanical data}

	\subsection{General description of study and variables}
	
%
%

		In order to assess biomechanical changes in typical training sessions in recreational runners on an equal level, $20$ participants ($10$ women and $10$ men) were initially selected to complete four typical training sessions sufficiently spaced in time. 	 Two were high-intensity interval training, and the rest were continuous training. In the first case, athletes ran $6\times 800(m)$ intervals at 1 km/h below their maximum aerobic speed with $1:1$ recovery. While in the second one, the athletes completed a continuous at a speed below maximum steady stater. The duration of the continuous run was individualised to the same estimated energy expenditure as the interval training. In addition, the training sessions were conducted at the same time of day to avoid possible daytime fluctuations.

	Running kinematics was measure with three-Dimensional motion analysis system that collect data at frequency of 500 Hz. All sessions were performed in an environmentally controlled laboratory setting, the athletes all used the same treadmill. Isometric strength was for various actions were recorded pre and post run. 
	
\begin{figure}[ht!]
		\centering
		\includegraphics[width=0.7\linewidth]{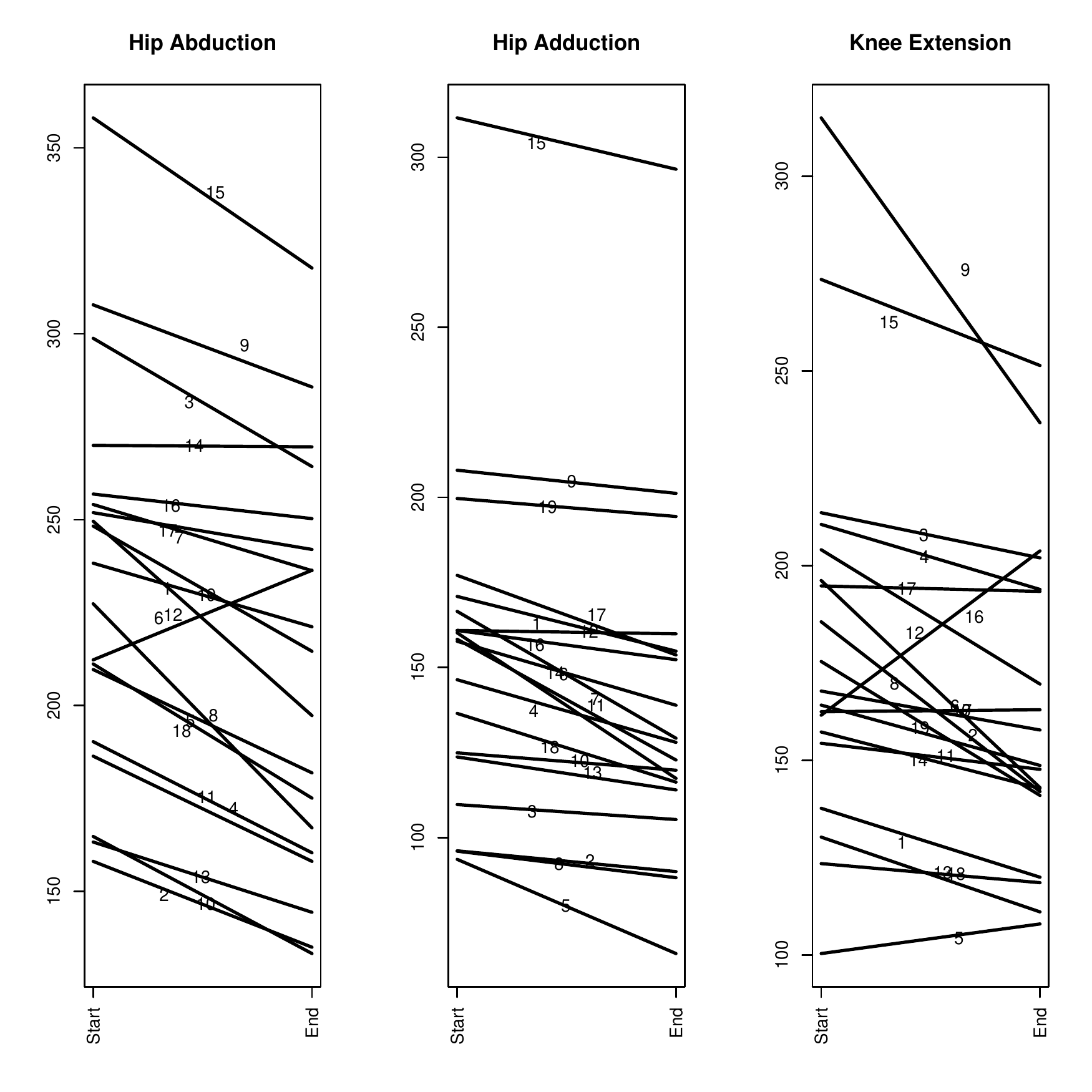}
		\caption{Changes in strength production at the start and end of last HIIT session for hip abduction, hip adduction and knee extension.}
		\label{fig:fuerza1}
	\end{figure}

	The participants' basic characteristics can be found in Table \ref{table:tabla2}. The strength changes in the last 800-m interval  in Hip Abduction, Hip Adduction, Knee Extension are shown in Figure \ref{fig:fuerza1}.

	In this paper, we analyze $20$ cycles of the stance phase for each run over $19$ participants. For security reasons, we have excluded one of the participants, due to the presence of some outliers and missing data in some part of strides.

	In our analysis, we have only focused on what happens in the three-dimensional knee segments: Knee-X, Knee-Y and Knee-Z. Additional details, about the study design and how the measurements were made can be found for example in \cite{10.3389/fbioe.2020.00360}.
	
%
%
%

	\begin{table}[ht!]
	\begin{center}
	\begin{tabular}{|c|c|c|}
		\hline
	Variable	& Female & Male  \\
		\hline
	Age (years)	& $42.3 \pm 4.4$  &  $43.8 \pm 4$  \\
		\hline
	Height (cm)	&  $164.8 \pm 6.3$  & $181.2 \pm 7.9$  \\
		\hline
	Mass (kg)  & $58.3 \pm 6.6$	&  $77.3\pm 6.5$    \\
		\hline
	HIIT Speed ($m\cdot s^{-1}$)	&  $3.9 \pm 0.3$  & $4.6 \pm 0.3$  \\
		\hline 
	HIIT rep duration (min:sec)   	& $3:24 \pm 13(s)$  & $2:47 \pm 16(s)$  \\
		\hline
	MICR Speed $(m\cdot s^{-1})$	& $3.3 \pm 0.2$ & $3.6 \pm 0.4$  \\
		\hline
	MICR duration (min:sec)            	& $32:16 \pm 2:03$ & $25:53 \pm 3:40 $\\
		\hline
	  $V0_{2}$ max ($ml\cdot kg^{-1}\cdot min^{-1}$)	& $52.8 \pm 5.0$ & $60.5\pm 4.4$ \\
		\hline
	sLTP ($m\cdot s^{-1}$)	& $3.4 \pm 0.1$  & $3.9 \pm 0.3$ \\
		\hline
	$\%$ $V0_{2}$ max at sLTP & $81.3 \pm 5.9$	 & $72.7 \pm 8.1$ \\
		
		\hline
	\end{tabular}
\caption{Descriptive characteristics of participants, training runs, speeds, durations,  $V0_{2}$ max, speed at lactate turnpoint (sLTP), percentage of $\%$ $V0_{2}$ max at sLTP ($\%$ $V0_{2}$ max at sLTP), represented as mean $\pm$ standard deviation.}
\label{table:tabla2}

		\end{center}
	\end{table}

	\subsection{Aims of the analysis, clinical implications and statistical analysis}

Knee injuries are the most common injury among runners of all levels. Therefore analyzing changes in posture and differences during typical training sessions has high clinical value in acquiring new epidemiological knowledge related to the causes of injuries. Examining the reliability between two interval training sessions is of fundamental importance. This will enable us to know how much information needs to be captured in order to characterize an athlete's biomechanical profile in a training session . In addition, changes in biomechanical patterns are very individualized and variable between individuals, therefore, using statistical tools that put the focus on the average subject in the study population rather than on individual variations can be very misleading, particularly in studies where the sample size is minimal. Finally, it is of interest to know if the information registered through the functional profiles and analyzed by the multilevel models provided information of the changes that occur during the athletes’ training sessions.

	In order to answer some of the questions mentioned above, we have divided the statistical analysis made with the three-dimensional information of the knee, in the following items:
	
	\begin{enumerate}
		\item  Examine the correlation between the scores obtained after applying the functional component analysis with the force production changes in the training session. 	
		\item  To estimate the multilevel functional intraclass correlation coefficient to measure the reliability between two interval training sessions using the 20-step information.
		
		\item To establish if statistically significant differences exist between a continuous and an interval training session by means of a hypothesis test exploiting the representation constructed from a functional multilevel model.		
\end{enumerate}
	
	In particular, we have selected the three-level nested model (N3) from the Table \ref{table:sec1} to carry out the model of all the previous issues.

	In the analyses outlined, the results will be accompanied by graphs that help us understand and discuss how individualized the biomechanical changes are and how it is a useful hypothesis test to infer conclusions in this context.

	\section{Results}

\begin{figure}[ht!]
			\subfloat[fig 1]{\includegraphics[width = 2.5in]{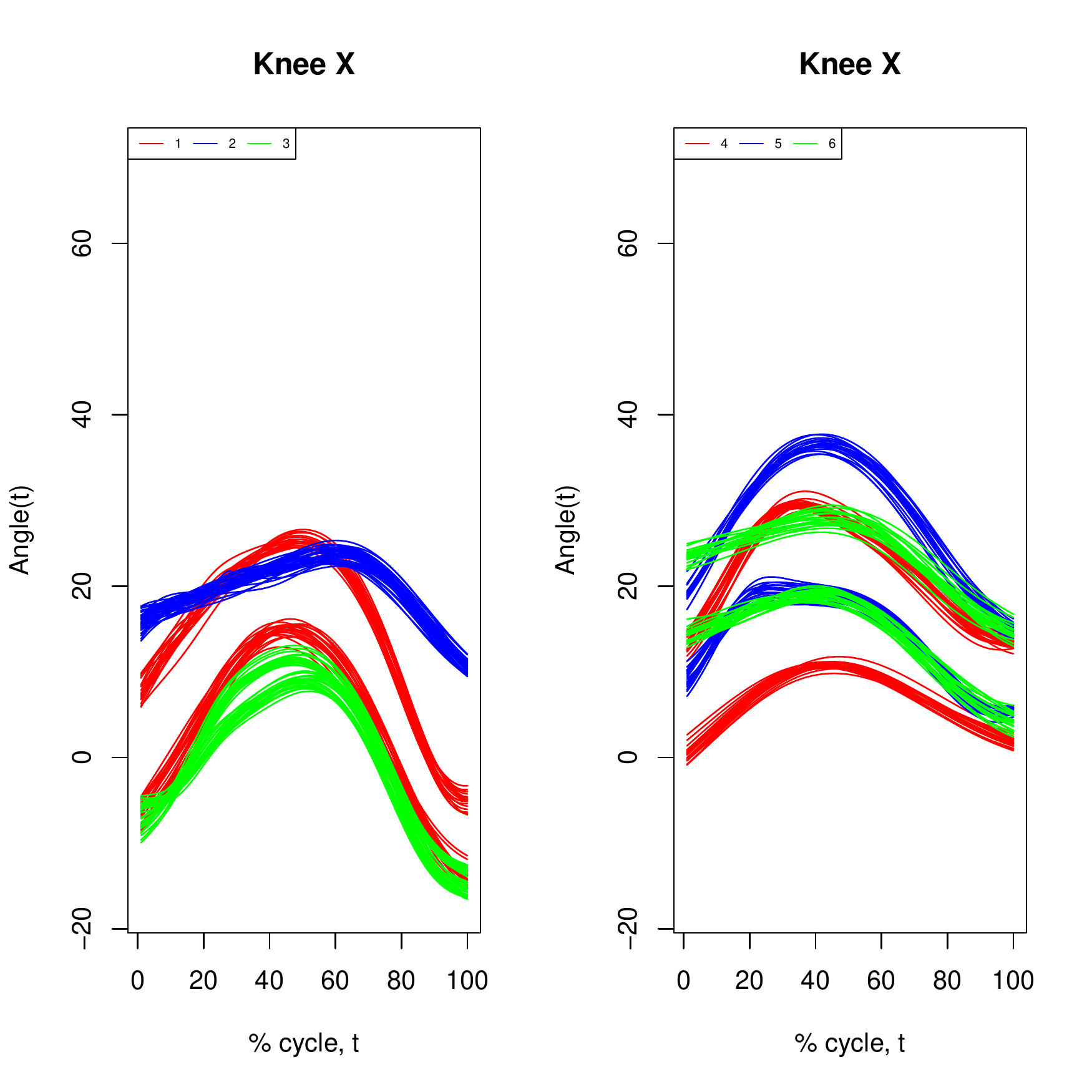}} 
			\subfloat[fig 2]{\includegraphics[width = 2.5in]{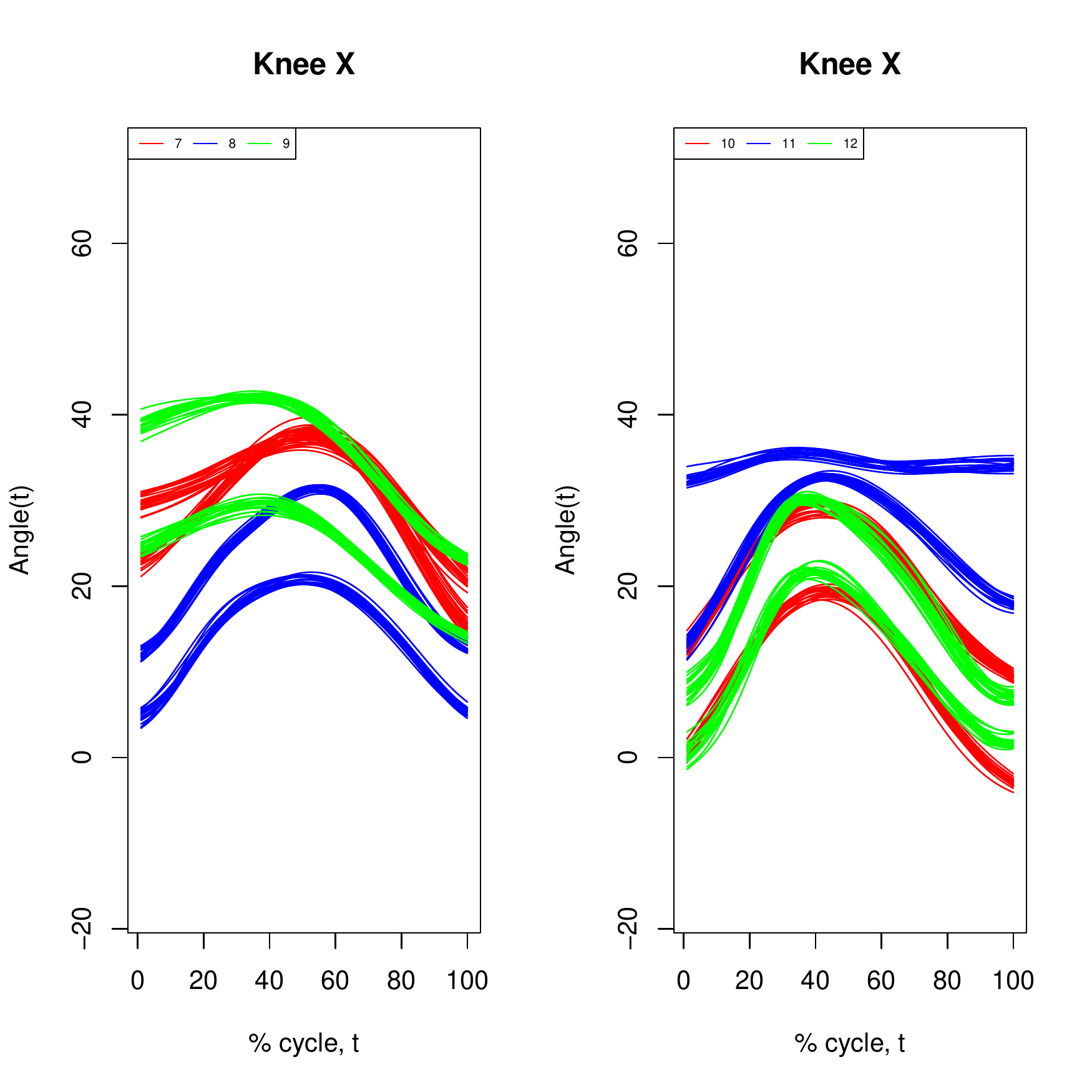}}\\
			\subfloat[fig 3]{\includegraphics[width = 2.5in]{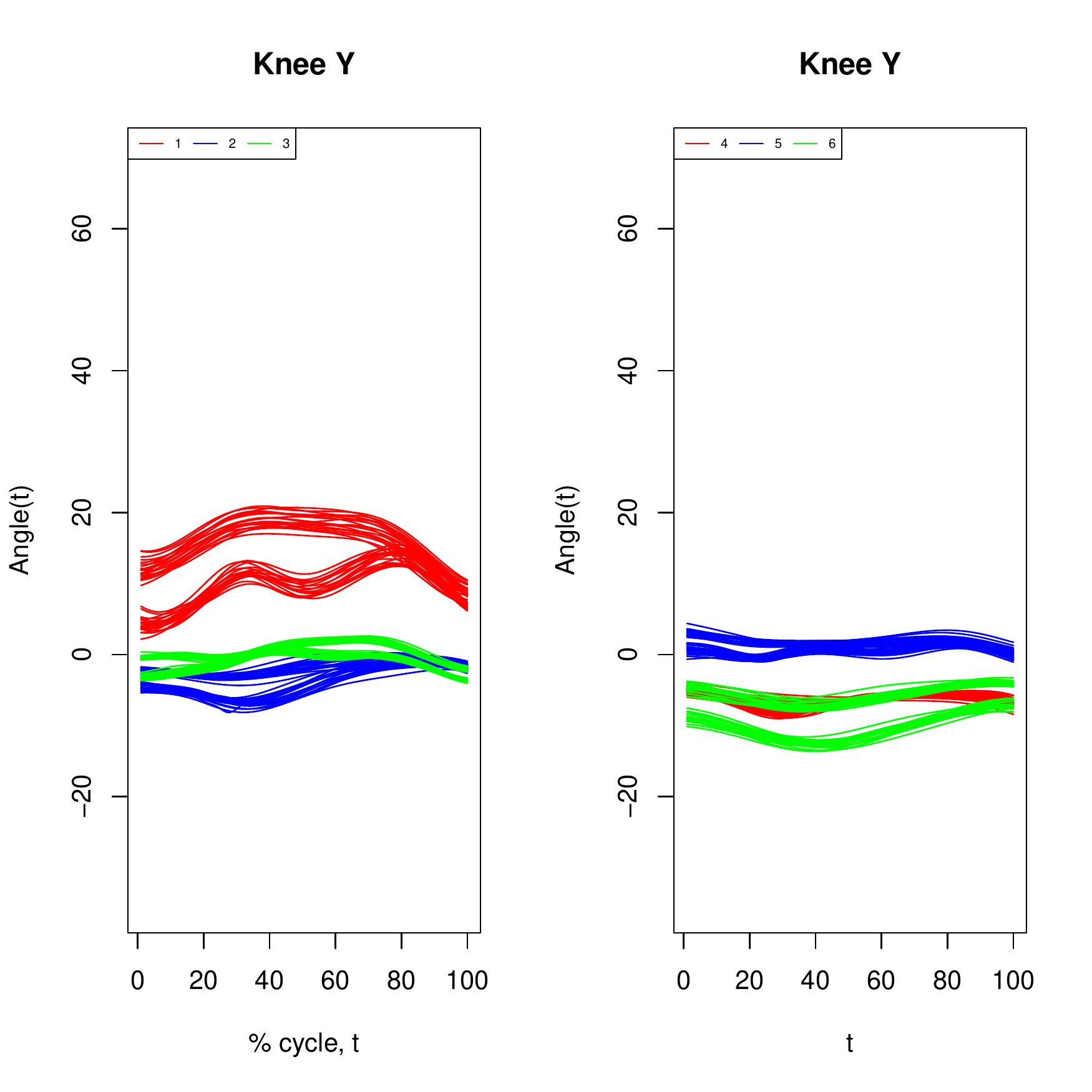}}
			\subfloat[fig 4]{\includegraphics[width = 2.5in]{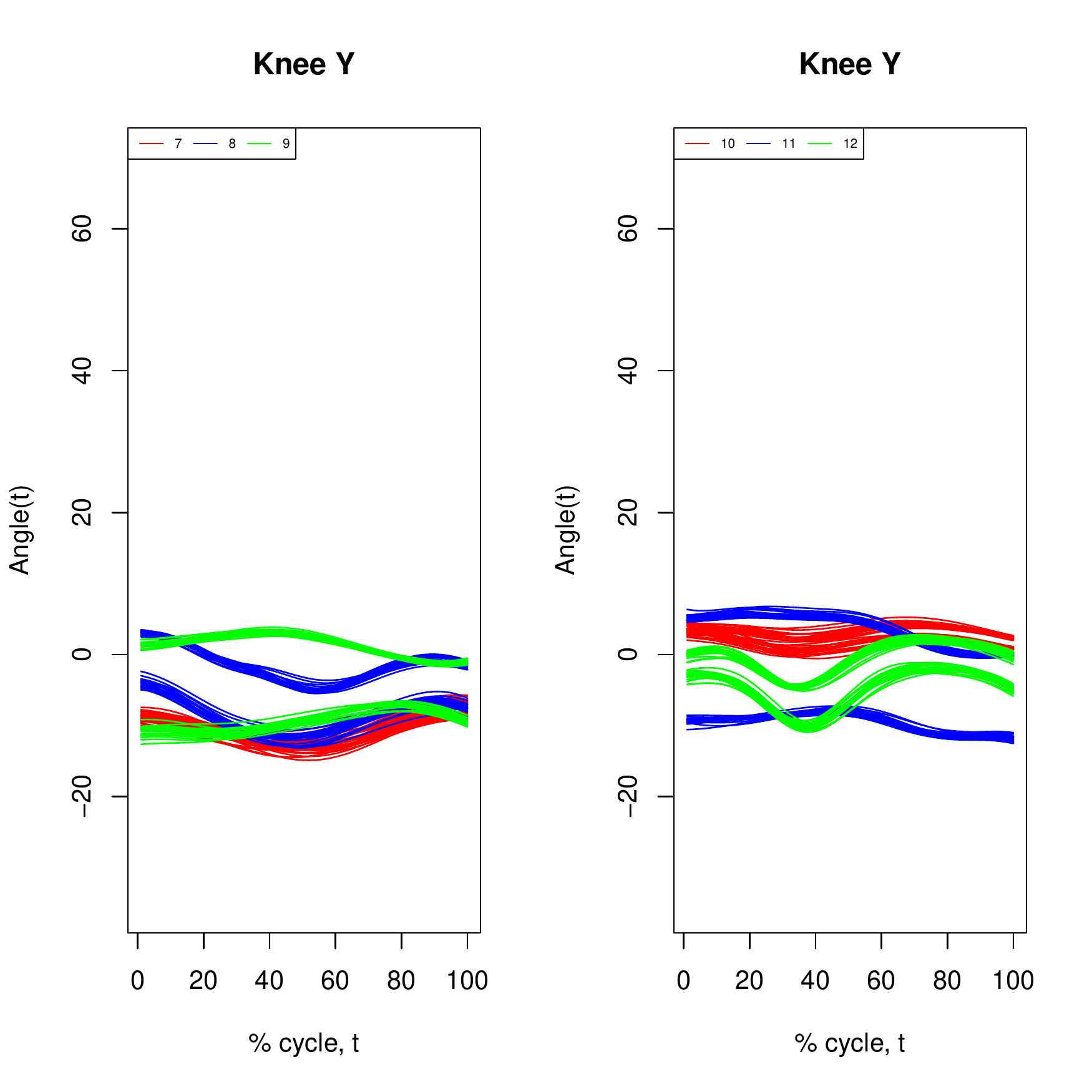}}\\
			\subfloat[fig 5]{\includegraphics[width = 2.5in]{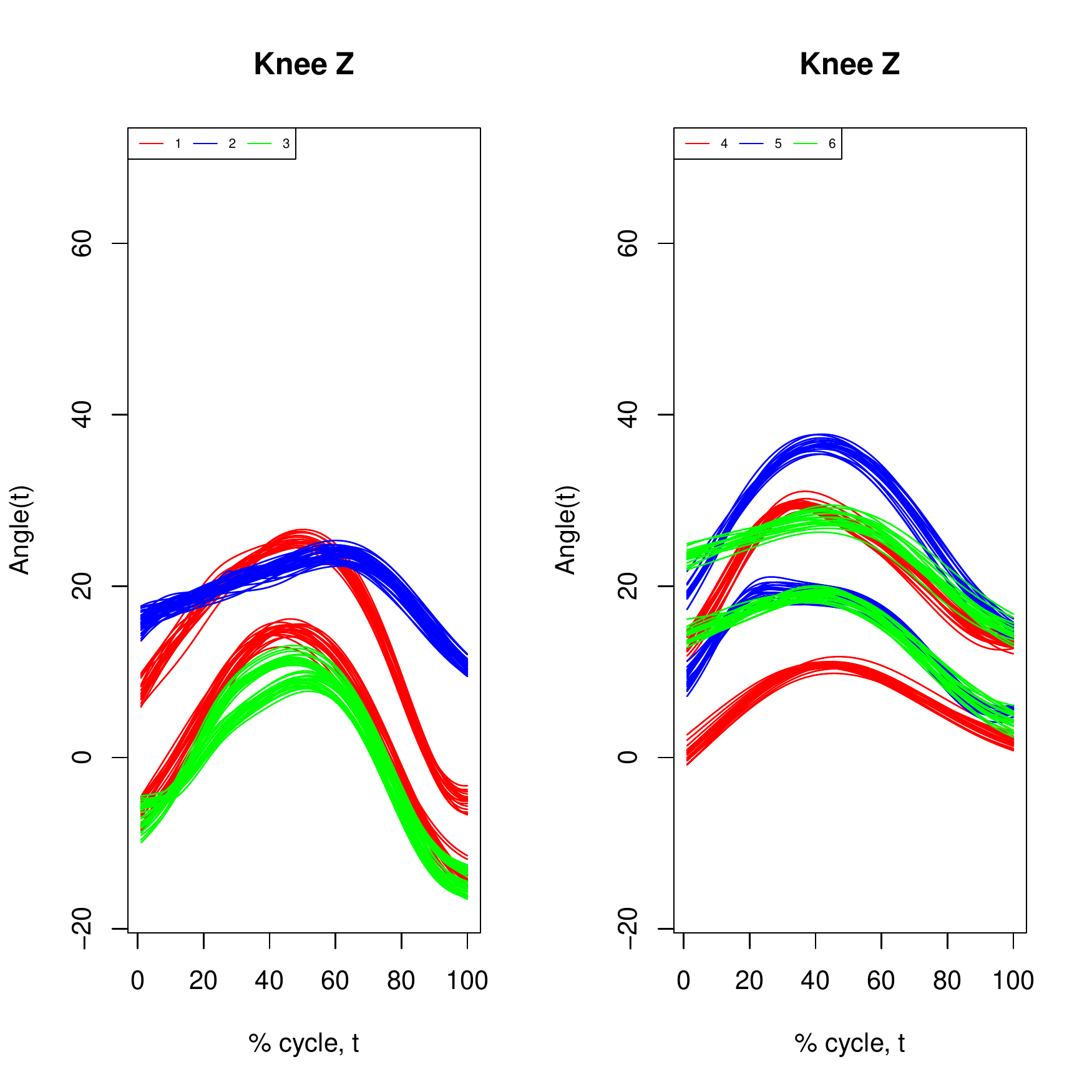}}
			\subfloat[fig 6]{\includegraphics[width = 2.5in]{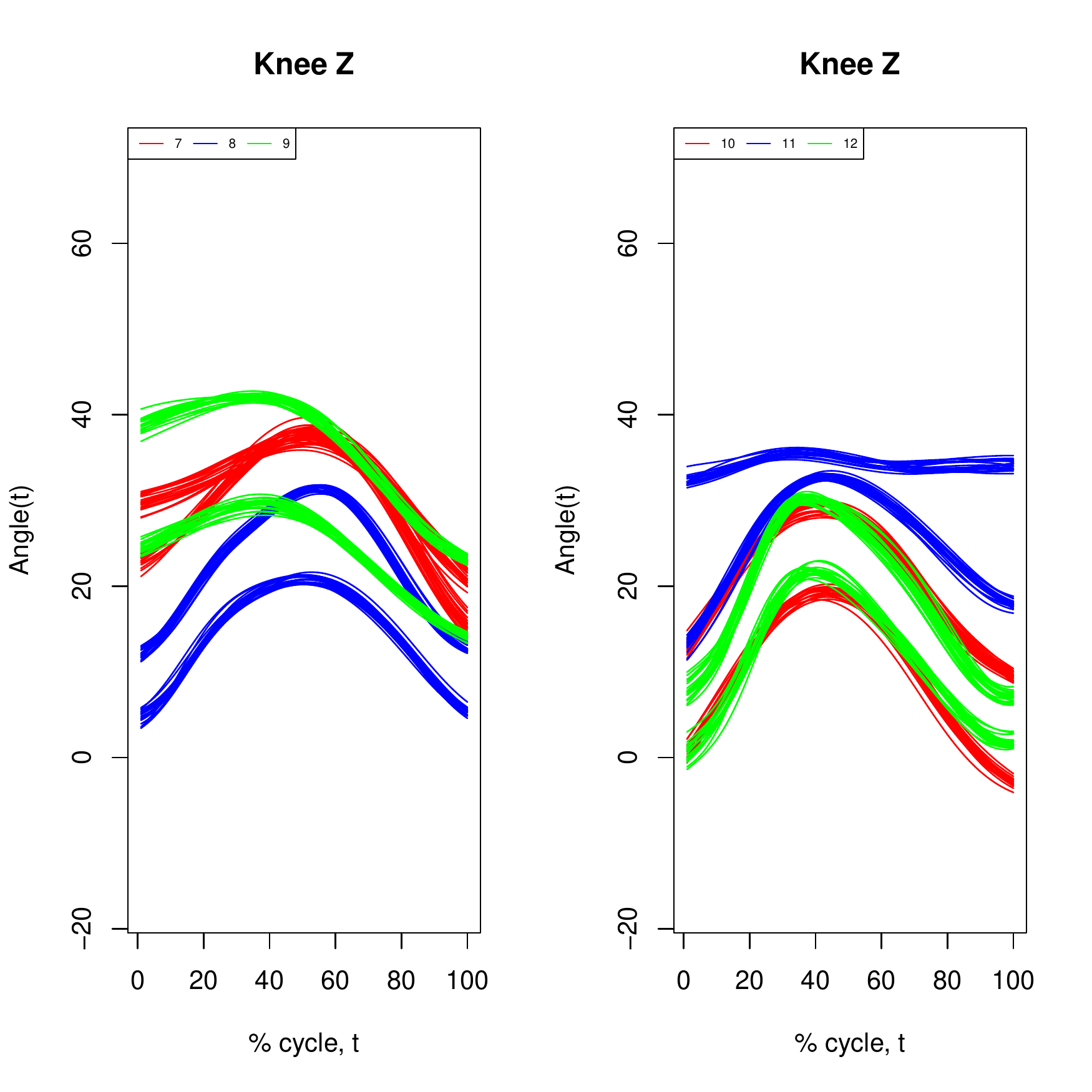}}\\
			\caption{Angle profiles were recorded along $20$ strides in two  HIIT sessions. Each individual in the same plot has their curves in the same color. We show graphics for Knee-X, Knee-Y, and Knee-Z. }
			\label{fig:atletas1}
		\end{figure}

\begin{figure}[ht!]
		\subfloat[fig 1]{\includegraphics[width = 2.2in]{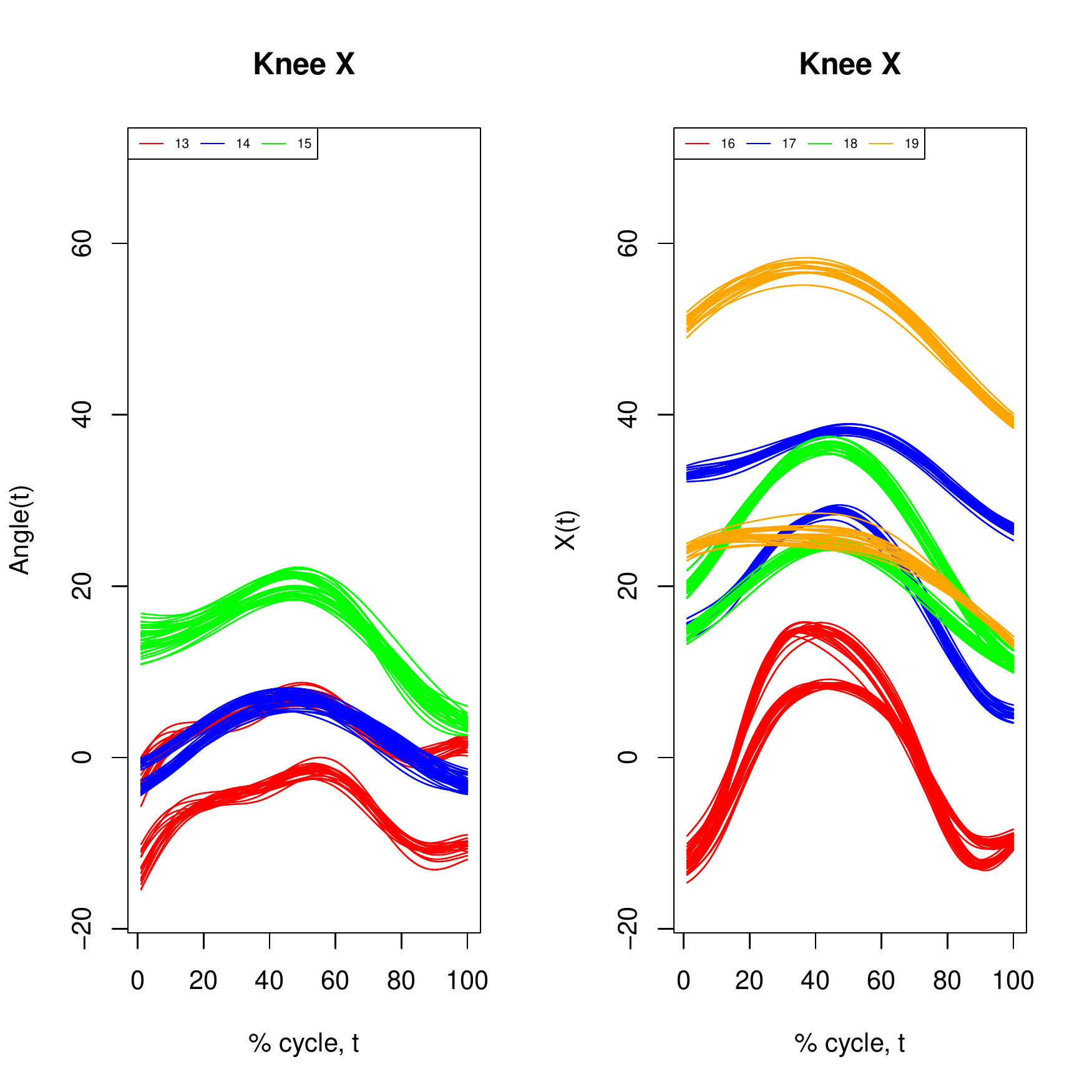}}		
	\subfloat[fig 2]{\includegraphics[width = 2.2in]{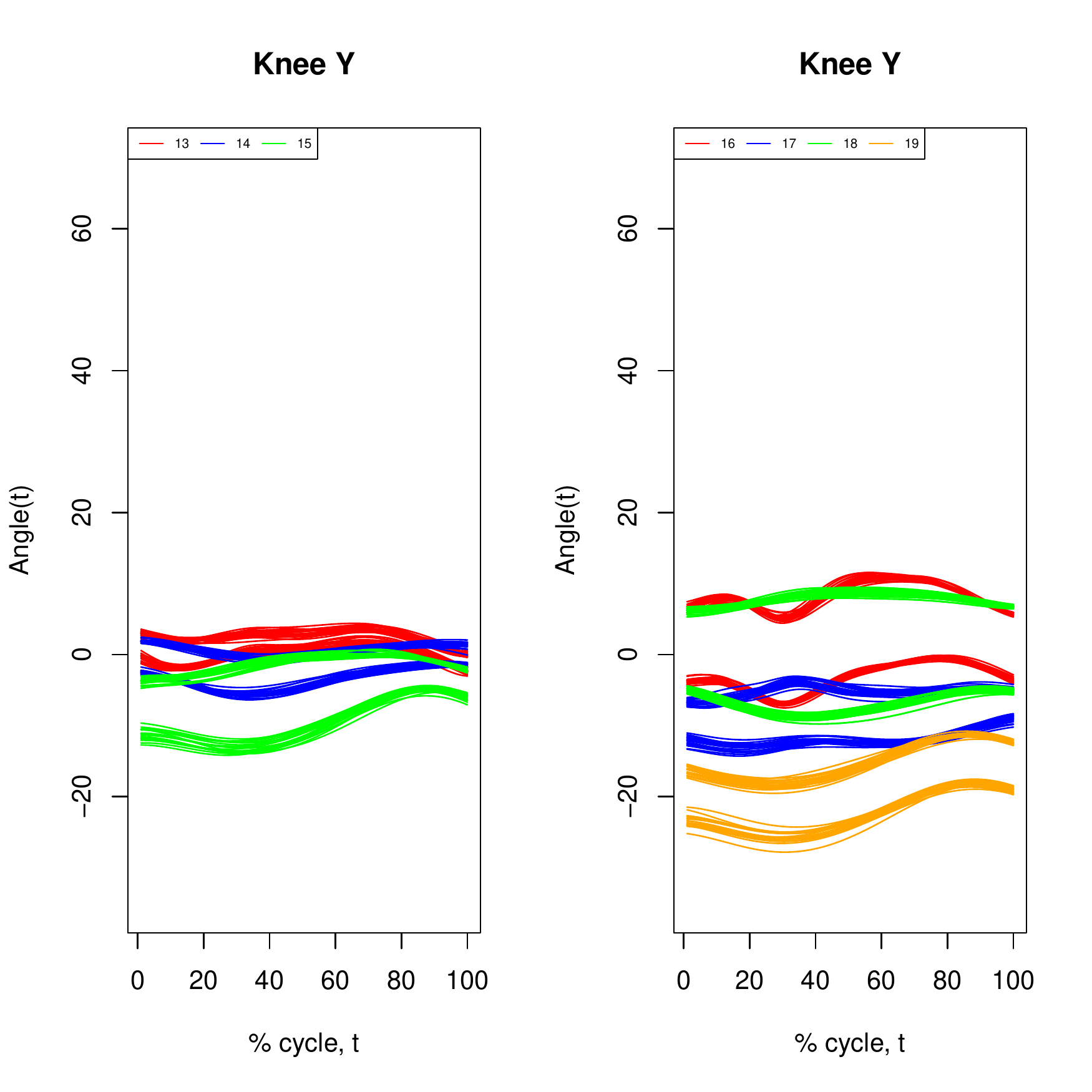}}\\
	\begin{center}
	\subfloat[fig 3]{\includegraphics[width = 2.2in]{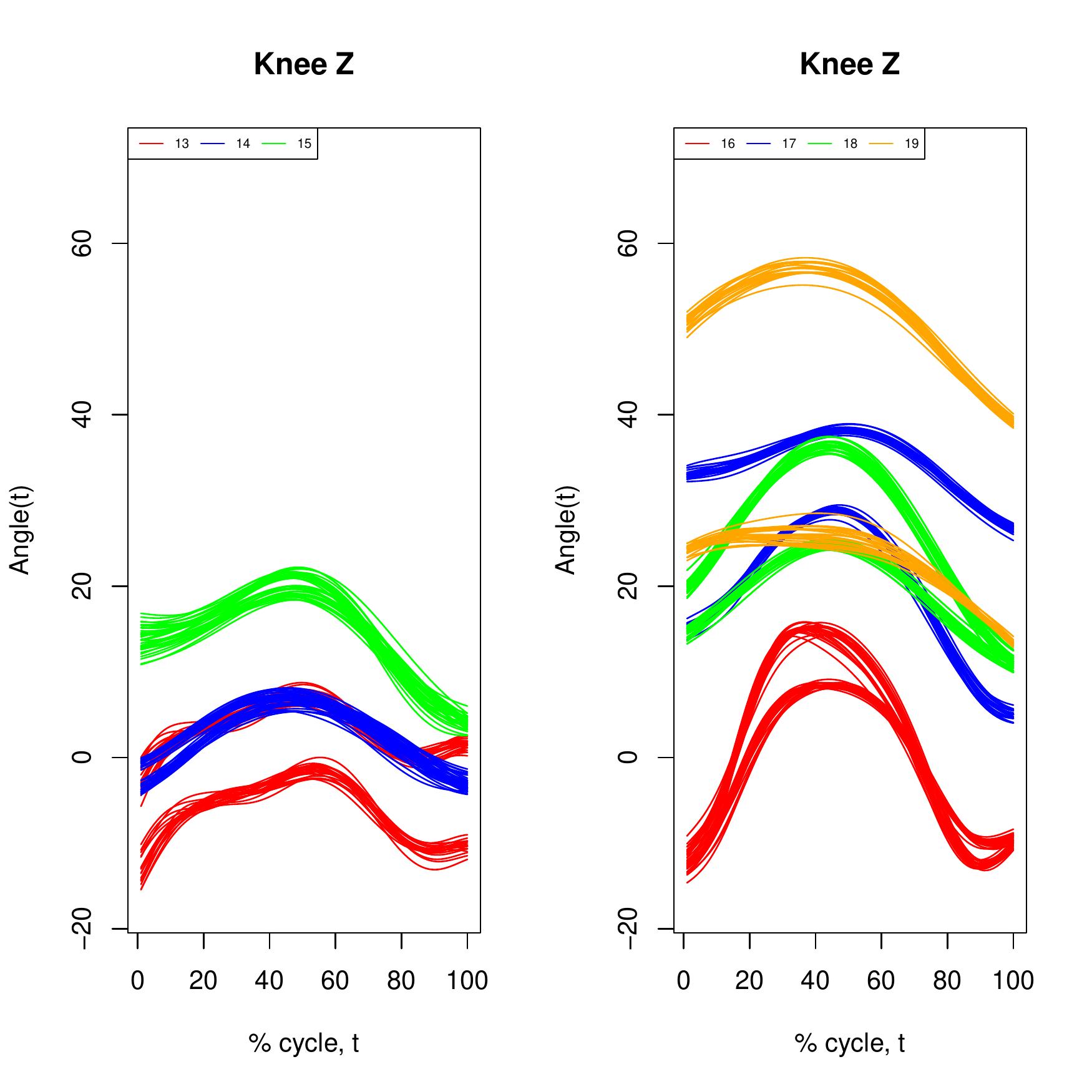}}
	\end{center}

		\caption{Angle profiles were recorded along $20$ strides in two  HIIT sessions. Each individual in the same plot has their curves in the same color. We show graphics for Knee-X, Knee-Y, and Knee-Z.}
		\label{fig:atletas2}
	\end{figure}

	Figures \ref{fig:atletas1} and \ref{fig:atletas2} contain the information about 20 strides per individual in different two HIIT runs in Knee-X, Knee-Y and Knee-Z. We can see that there are subjects in whom there are hardly any differences in their biomechanical profiles between the two runs. However, in others,  differences seem to be present. In addition, we can also see that the patterns between the two runs are quite individual; no common pattern in angle values exists across the all the runners examined.

\begin{figure}[ht!]
	\subfloat[fig 1]{\includegraphics[width = 5.0in, height=2.5in, scale=0.6]{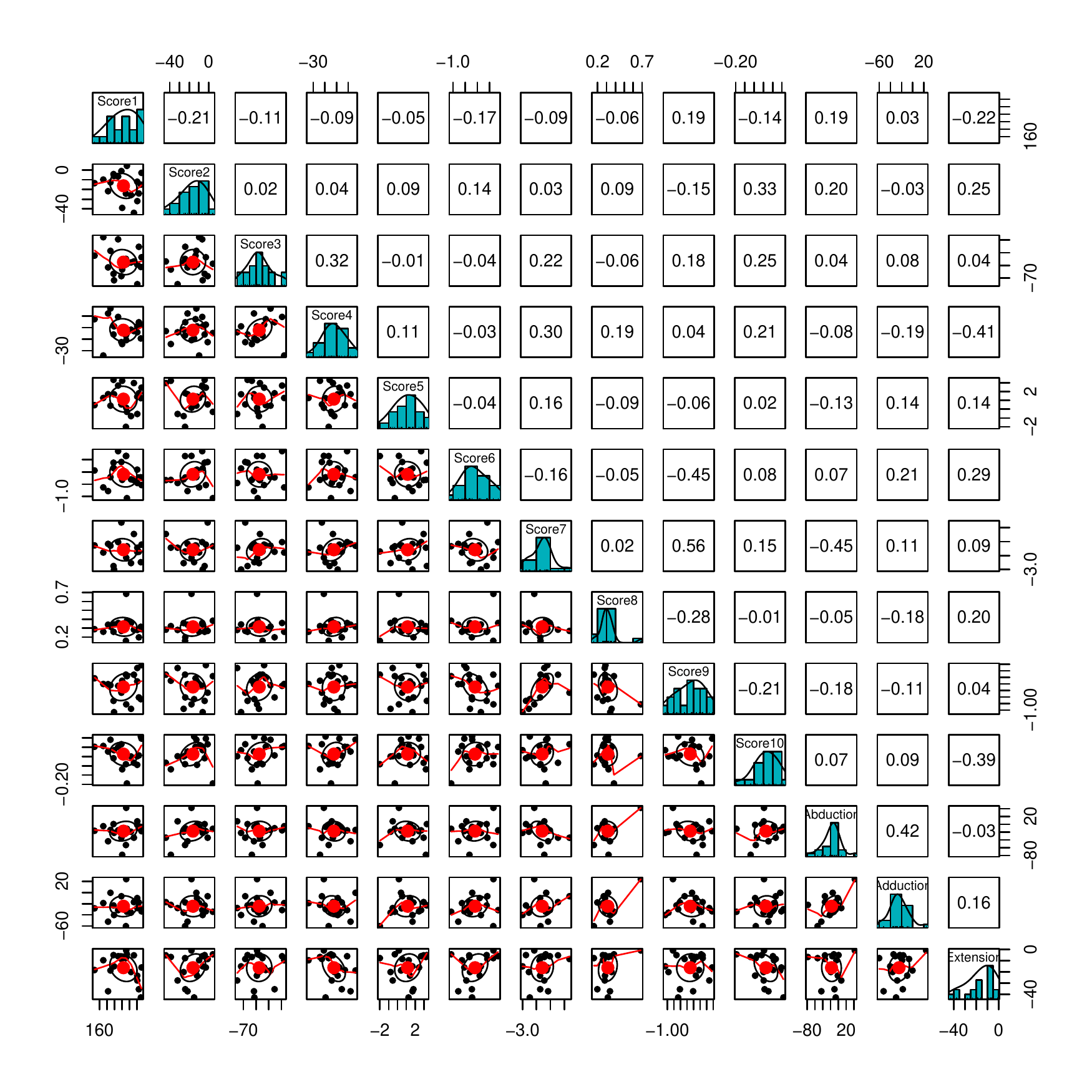}}\\ 
	\subfloat[fig 2]{\includegraphics[width = 5.0in, height=2.5in, scale=0.6]{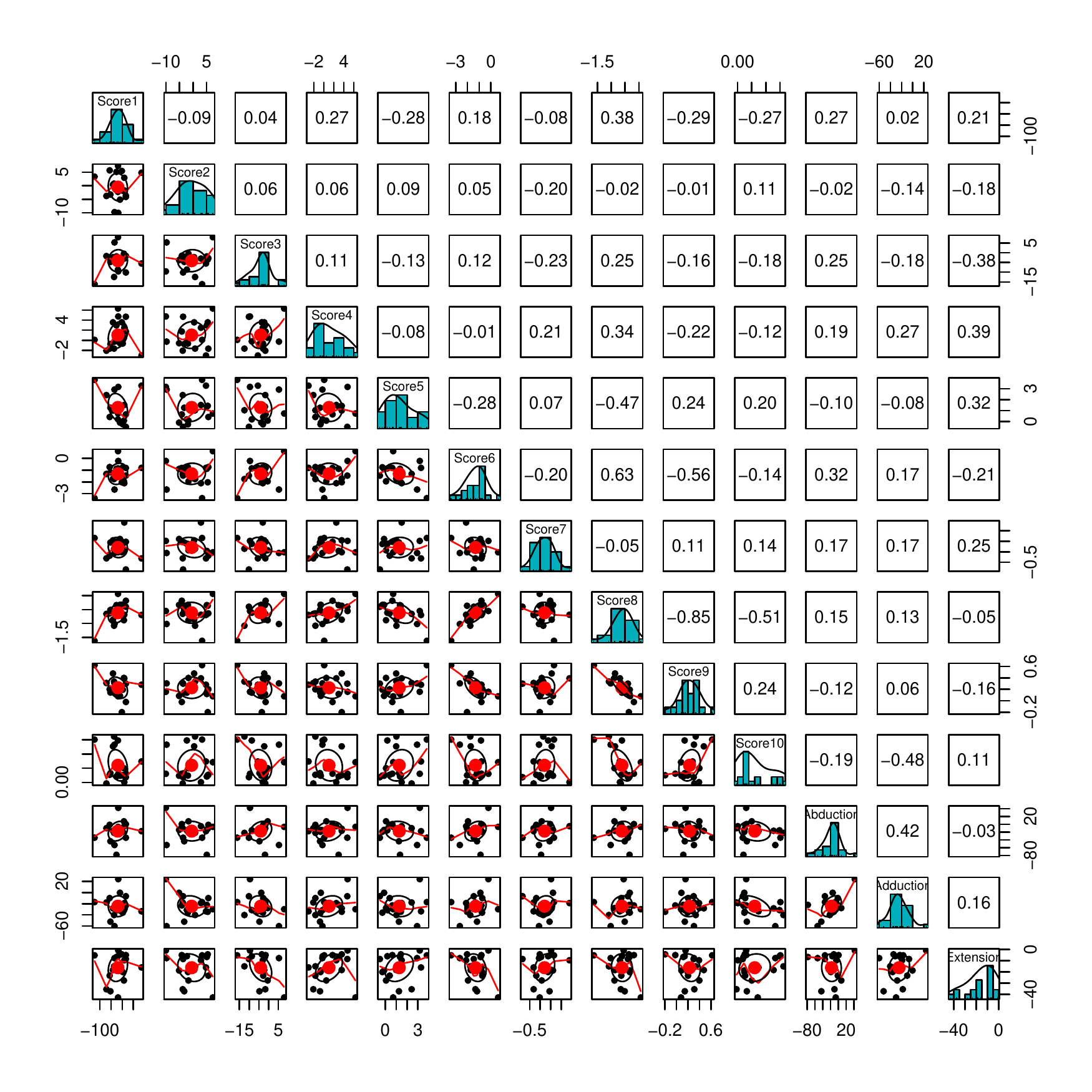}}\\
	\subfloat[fig 3]{\includegraphics[width = 5.0in, height=2.5in, scale=0.6]{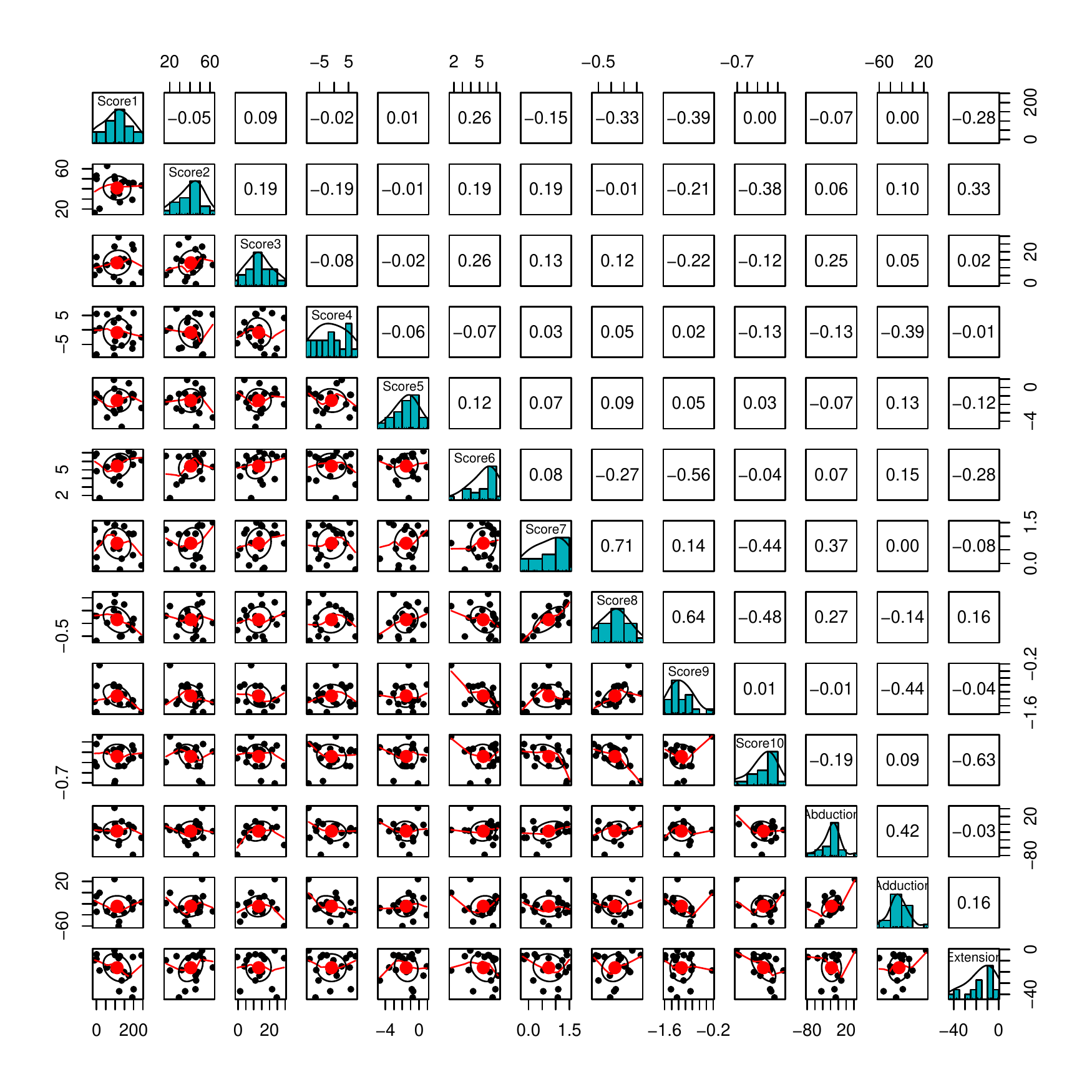}}\\
	\caption{ Spearman-correlation and bidimensional plots between functional scores calculated with multilevel models  and change in strength production in Hip Abduction, Hip Adduction and Knee Extension}
	\label{fig:correlacion1}
\end{figure}

	Figure \ref{fig:correlacion1} show the bivariate association between each functional scores after applied multilevel principal component analysis and changes in the strength production between athletes. The results show that in some scores, there is a significant correlation against changes in strength production. However, in other cases, the correction is very poor. Notwithstanding this, we examined the marginal association and probability can be the interaction in more complex models between scores, but the limited sample size of this study remains fit more complex model.

	The functional ICC for Knee-X is $0.55$, Knee-Y $0.54$, and  Knee-Z  $0.61$.
	
\begin{figure}
		\subfloat[fig 1]{\includegraphics[width = 2.5in]{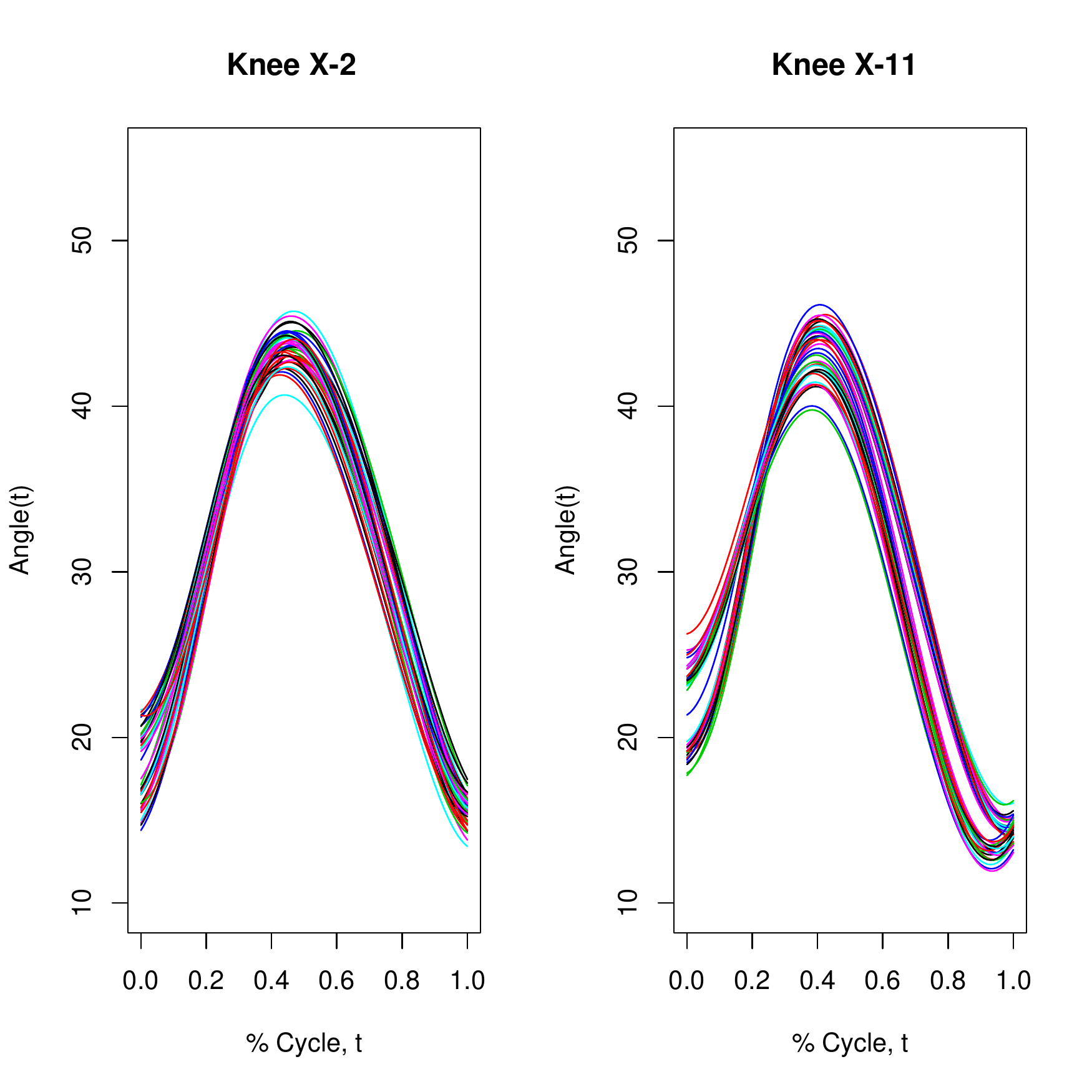}}		
			\subfloat[fig 3]{\includegraphics[width = 2.5in]{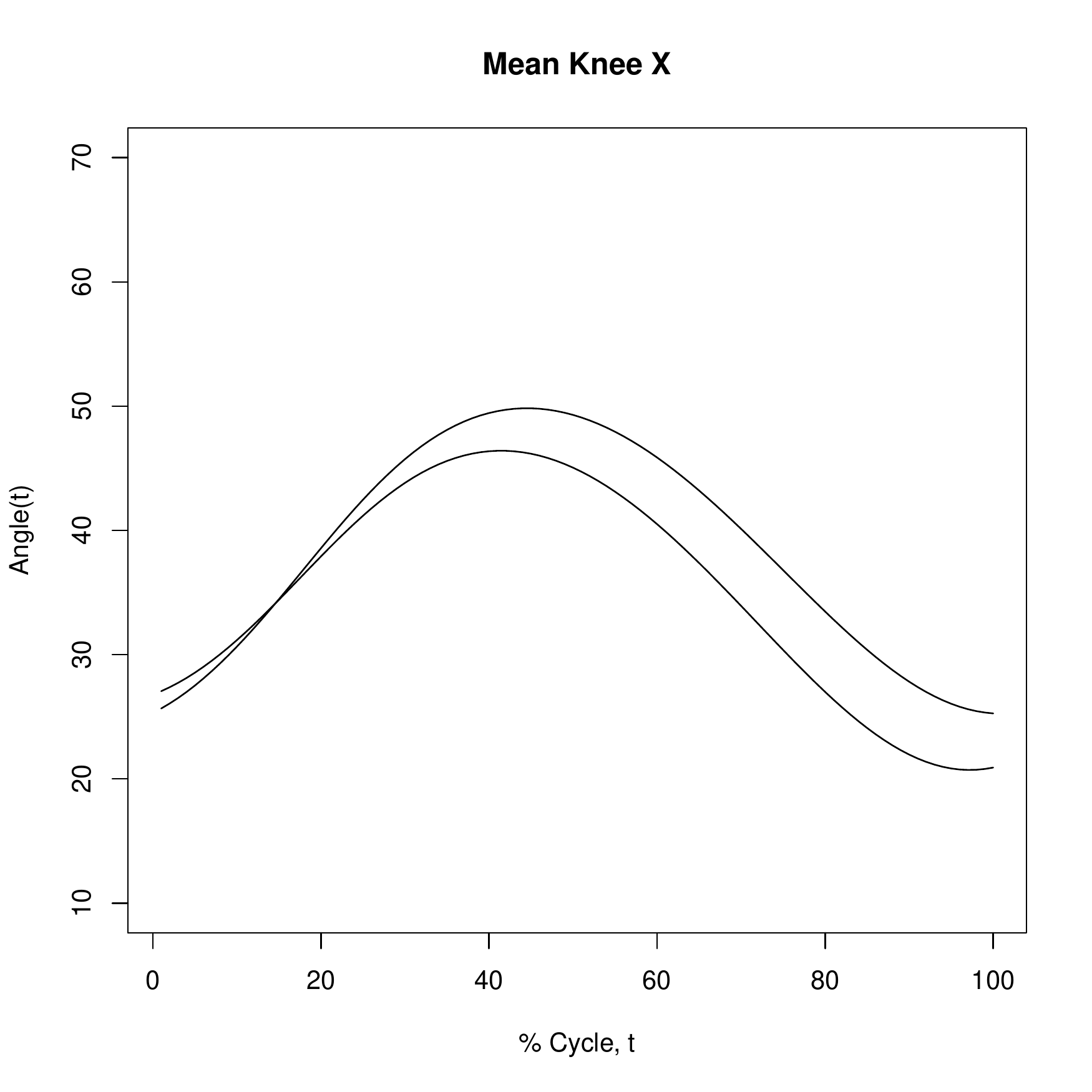}}
		\caption{Biomechanical pattern of two athletes between HIIT session and continuous running and mean functional curve in each run for all runners}
		\label{fig:cambios1}
	\end{figure}

\begin{figure}
		\subfloat[fig 1]{\includegraphics[width = 2.5in]{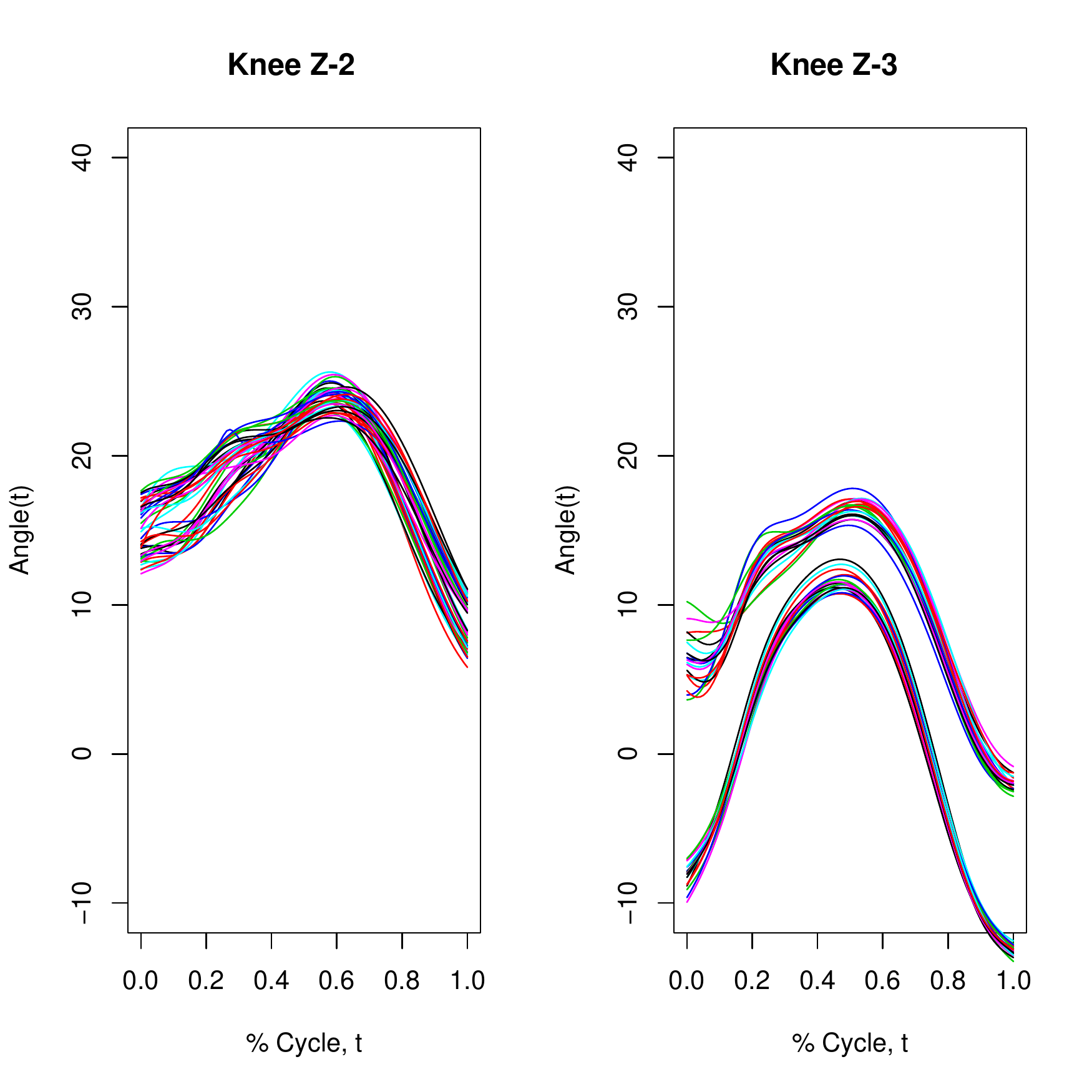}}		
		\subfloat[fig 3]{\includegraphics[width = 2.5in]{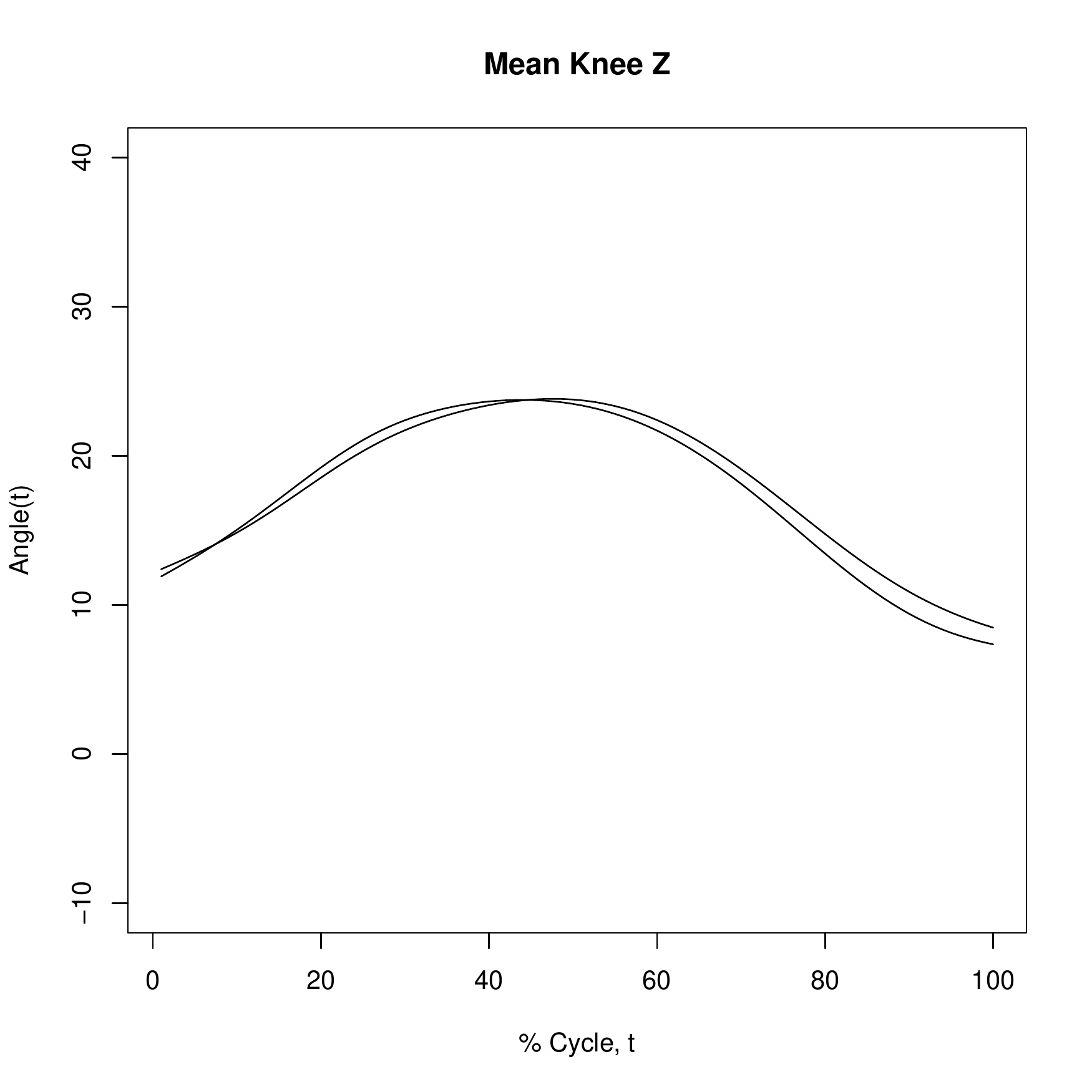}}
		\caption{Biomechanical pattern of two athletes between HIIT session and continuous running and mean functional curve in each run for all runners}
		\label{fig:cambios3}
	\end{figure}

	Likewise, at a significance level of 5\%, no statistical differences were found between the biomechanical patterns during interval training and continuous running, with the p-values for Knee-X, Knee-Y, and Knee-Z respectively of $0.17$, $0.12$ and $0.4$.  Figures \ref{fig:cambios1}, \ref{fig:cambios3} shows the measured curves of each group taken with the average of the $20$ steps, along with the biomechanical profiles of two athletes we consider representative. They show that the biomechanical changes between an interval run and a continuous run among the athletes are very changeable. In some individuals, there is no biomechanical changes in running style, while in others, the biomechanic profiles are very different. However, the mean values are not significantly different.

	\section{Discussion}

	Knee injuries are one of the most frequent problems faced by recreational runners \cite{van2007incidence}. An accurate characterization of the biomechanical changes that occur in typical training sessions can be critical in identifying the etiology of injuries \cite{donoghue2008functional} and developing predictive models to  detect injury risk \cite{ceyssens2019biomechanical}.  Here, we have illustrated how to exploit the functional information of different steps during different training sessions from multilevel models to:  i) examine the correlation between knee angles and changes in force production in the same training session; ii) measure the reliability between two training sessions: iii) see that there are no statistically significant differences between a continuous run and an interval training with the same energy expenditure, although remarkable differences exist if we visually analyze some individuals.

	 The complete analysis of each cycle through functional analysis techniques that analyze the curve in its totality has lead to more nuanced findings \cite{donoghue2008functional}. Traditional techniques that analyze either fixed angles, the average angle, the range of movement or other measures summarized, result in  the loss of information that its use entails. 
		Complementary,  interesting problems can be identified when using more informative gait points. Recent statistical methodologies can be used to address this problem \cite{berrendero2016variable,	poss2020superconsistent}

	Functional multilevel models are an essential weapon in the challenge to exploit information from monitoring athletes or patients, to optimize decision making using different sources of information and measurements, made at different resolution levels. These tools can help integrate and analyze the information together, obtain a representation of the individuals along with different levels of hierarchy, and establish the different forms of variability in the different levels considered. These tools are remarkable if we want to analyze all training records or physiological variables of a group of athletes over a season or different micro-macro-cyc*les \cite{lambert2010measuring,halson2014monitoring}. For example, there is not yet a sufficiently good methodology to represent the information inherently as proposed by these models \cite{matabuena2019improved, piatrikova2021monitoring, kalkhoven2021training}.   Despite being an exciting research topic with high relevance, we believe that there are not many methodologies to address relevant problems  in biomechanics to date. For example, a specific need of this field could be to build a multilevel model that considers the different time length of the step, and not lose information of the step geometry with the standardization of all the strides to the $[0,1]$ interval.

	The multilevel models have allowed us to calculate the intraclass correlation coefficient between the two interval training sessions taking into account the $20$ steps recorded in each  session. To the best of our knowledge, this is a novel approach in this area since the traditional approaches previously used to measure reliability rely on the compression of information in the average curve and only beetween two conditions \cite{pini2019test}. At the same time, we have introduced a new hypothesis test to test the statistical differences between continuous and multilevel running, taking advantage of the representation we obtained with the multilevel model at the second level of the hierarchy. This also represents an advance, since with the inclusion of the $20$ steps in the model in each test, we have more information, and with the new procedure, we can see if there are statistical differences between the different levels of hierarchy or groups of patients/athletes taking into account the differences in the study design.

An important aspect to consider in analyzing the results is that the individuals' movement patterns seem unique. This is not new, and several papers have exempted the individuality of human walking and running \cite{horst2019explaining}. In this sense, since the biomechanical patterns are probably grouped in clusters \cite{phinyomark2015kinematic, jauhiainen2020hierarchical}, standard hypothesis tests applied to the whole sample are not the best way to establish biomechanical differences. There are some discrepancies between studies when examining these issues. Also, in the biomechanics literature, as in other biomedical literature areas, there is some controversy about the use of p-value \cite{benjamin2018redefine}, and the use of other approaches such as effect size \cite{browne2010t} or e-values \cite{vovk2019values} may be recommended.

A limitation of this study is the sample size, together with the fact that we are analyzing the biomechanical variations of the knee, without taking into account the possible multivariate structure of knee movement. However, due to the reduced number of data, we can gain a greater interpretation in this type of study of a more exploratory character with this procedure. Moreover, this work's main objective is to illustrate the use of multilevel models with biomechanical data.

The rise of biosensors \cite{ferber2016gait,phinyomark2018analysis} in the area of biomechanics and medicine is causing an unprecedented revolution in the evaluation of athletes and patients care. It is likely that in the coming years, many of the clinical decisions will also be supported by the values predicted from the algorithms in many contexts, such as the prediction of  injuries \cite{clermont2020runners, van2020real} or optimal surgery recovery \cite{karas2020predicting, kowalski2021recovery} so in sport and general populations. Undoubtedly, the introduction of the data analysis techniques discussed here will help practitioners analyze objects that vary in a continuum repeatedly and that appear more and more frequently in biomedical data \cite{dunn2018wearables}.

 \section*{Acknowledgements}

Marcos Matabuena thanks  Ciprian Crainiceanu for their email answers to specific questions related to functional multilevel model methodology developed for their research group in the last two decades.

 This work has received financial support from the Spanish Ministry of Science, Innovation and Universities under Grant RTI2018-099646-B-I00, the Consellería de Educación, Universidade e Formación Profesional and the European Regional Development Fund under Grant ED431G-2019/04.

\section*{Competing Interests}
 The authors declare no competing interests.
 
 \section*{ETHICS STATEMENT}

The studies involving human participants were reviewed and
approved by Northumbria University. The patients/participants
provided their written informed consent to participate in this study.

 \bibliographystyle{apalike}
 \bibliography{biblio}
	

\end{document}